\documentclass[fleqn,usenatbib]{mnras}

\usepackage{newtxtext,newtxmath}
 
\usepackage{graphicx}
\usepackage{makecell}
\usepackage{amsmath}
\usepackage{bm}
\usepackage{gensymb}
\usepackage{natbib}
\usepackage[T1]{fontenc}
\usepackage{ae,aecompl}

\title[WASP-127b transmission spectrum]{Abundance measurements of H$_{2}$O and carbon-bearing species in the atmosphere of WASP-127b confirm its super-solar metallicity} % 

\author[J. J. Spake et al.]
{Jessica J. Spake,$^{1,2,3}$\thanks{51 Pegasi b Fellow, E-mail: jessica.spake@gmail.com},
David K. Sing$^{1,2,4}$,         % Re-arrange as needed
Hannah R. Wakeford$^{5}$,
Nikolay Nikolov$^{6}$,
\newauthor
Thomas Mikal-Evans$^{7}$,
Drake Deming$^{8}$, 
Joanna K. Barstow$^{9}$,
David R. Anderson$^{10,11}$, 
\newauthor
Aarynn L. Carter$^{1}$,
Michael Gillon$^{12}$,
Jayesh M. Goyal$^{13}$,
Guillaume Hebrard$^{14,15}$,
\newauthor
Coel Hellier$^{10}$,
Tiffany Kataria$^{16}$,
Kristine W. F. Lam$^{17}$,
A. H. M. J. Triaud$^{18}$,
\newauthor
Peter J. Wheatley$^{11}$
\\
$^{1}$Physics and Astronomy, Stocker Road, University of Exeter, Exeter, EX4 3RF, UK\\
$^{2}$Department of Earth \& Planetary Sciences, Johns Hopkins University, Baltimore, MD, USA\\
$^{3}$Division of Geological and Planetary Sciences, California Institute of Technology, 1200 East California Blvd, Pasadena, CA 91125, USA\\
$^{4}$Department of Physics \& Astronomy, Johns Hopkins University, Baltimore, MD, USA\\
$^{5}$School of Physics, University of Bristol, HH Wills Physics Laboratory, Tyndall Avenue, Bristol BS8 1TL, UK\\
$^{6}$Space Telescope Science Institute, 3700 San Martin Dr, Baltimore, MD 21218, USA\\
$^{7}$Kavli Institute for Astrophysics and Space Research, Massachusetts Institute of Technology, 77 Massachusetts Avenue, 37-241, \\ Cambridge, MA 02139, USA\\
$^{8}$Department of Astronomy, University of Maryland, College Park, MD, USA\\
$^{9}$School of Physical Sciences, The Open University, Walton Hall, Milton Keynes\\
$^{10}$Astrophysics Group, Keele University, Staffordshire ST5 5BG, UK\\
$^{11}$Department of Physics, University of Warwick, Gibbet Hill Road, Coventry CV4 7AL, UK\\
$^{12}$Space sciences, Technologies and Astrophysics Research (STAR) Institute, Unieversite de Liege, Liege 1, Belgium\\
$^{13}$ Department of Astronomy and Carl Sagan Institute, Cornell University, 122 Sciences Drive, Ithaca, NY 14853, USA\\
$^{14}$Sorbonne Universite, CNRS, UMR 7095, Institut d'Astrophysique de Paris,98 bis bd Arago,75014 Paris, Fr\\
$^{15}$Observatoire de Haute-Provence, CNRS, Universited Aix-Marseille, 04870 Saint-Michel-l' Observatoire, Fr\\
$^{16}$NASA Jet Propulsion Laboratory, 4800 Oak Grove Drive, Pasadena, CA 91109, USA\\
$^{17}$Center for Astronomy and Astrophysics, Technical University Berlin, Hardenbergstr. 36, 10623 Berlin, Germany\\
$^{18}$School of Physics and Astronomy, University of Birmingham, Edgbaston, Birmingham B15 2TT, UK\\
}

\date{Received \today; in original form \today}%Accepted XXX. 

\pubyear{2019}

\begin{document}
\label{firstpage}
\pagerange{\pageref{firstpage}--\pageref{lastpage}}
\maketitle

\begin{abstract}
The chemical abundances of exoplanet atmospheres may provide valuable information about the bulk compositions, formation pathways, and evolutionary histories of planets.  Exoplanets with large, relatively cloud-free atmospheres, and which orbit bright stars provide the best opportunities for accurate abundance measurements.  For this reason, we measured the transmission spectrum of the bright (V$\sim$10.2), large (1.37\,R$_{J}$), sub-Saturn mass (0.19\,M$_{J}$) exoplanet WASP-127b across the near-UV to near-infrared wavelength range (0.3--5\,$\mu$m), using the Hubble and Spitzer Space Telescopes.  Our results show a feature-rich transmission spectrum, with absorption from Na, H$_{2}$O, and CO$_{2}$, and wavelength-dependent scattering from small-particle condensates.  We ran two types of atmospheric retrieval models: one enforcing chemical equilibrium, and the other which fit the abundances freely.  Our retrieved abundances at chemical equilibrium for Na, O and C are all super-solar, with abundances relative to solar values of 
9$^{+15}_{-6}$,  16$^{+7}_{-5}$, and 26$^{+12}_{-9}$ respectively.  
%Note that the parent star WASP-127A has a slightly sub-solar metallicity of [Fe/H]=-0.18$\pm$0.0, \citep{2017A&A...599A...3L}.
Despite giving conflicting C/O ratios, both retrievals gave super-solar CO$_{2}$ volume mixing ratios, which adds to the likelihood that WASP-127b's bulk metallicity is super-solar, since CO$_{2}$ abundance is highly sensitive to atmospheric metallicity.  We detect water at a significance of 13.7 $\sigma$. Our detection of Na is in agreement with previous ground-based detections, though we find a much lower abundance, and we also do not find evidence for Li or K despite increased sensitivity. In the future, spectroscopy with JWST will be able to constrain WASP-127b's C/O ratio, and may reveal the formation history of this metal-enriched, highly observable exoplanet.

%This is a simple template for authors to write new MNRAS papers.
%The abstract should briefly describe the aims, methods, and main results of the paper.
%It should be a single paragraph not more than 250 words (200 words for Letters).
%No references should appear in the abstract.
\end{abstract}


%%%%%%%%%%%%%%%%%%%%%%%%%%%%%%%%%%%%%%%%%%%%%%%%%%

%%%%%%%%%%%%%%%%% BODY OF PAPER %%%%%%%%%%%%%%%%%%

\section{Introduction}
WASP-127b is a transiting, sub-Saturn mass exoplanet which was discovered by the SuperWASP survey \citep{2017A&A...599A...3L}. It has the largest expected atmospheric scale height of any planet yet discovered, at $\sim$2\,350km. Its host star is bright (V$\sim$10.2, J$\sim$9.1), and appears to be old, photometrically quiet, and slowly-rotating\footnote{ WASP-127's age estimate from isochrone fitting is 11.41 $\pm$ 1.80 Gyr; the SuperWASP photometry shows no sign of variability;  and the stellar $v$sin$i$ value is too small to measure from high-resolution spectra  \citep{2017A&A...599A...3L}}.  These planetary and stellar properties make WASP-127b a standout target for atmospheric characterization, because they result in large transmission signals which are not affected by stellar variability.  With the Hubble (HST) and Spitzer Space Telescopes, it is possible to measure a transmission spectrum rivaling the quality of even the canonical planets HD 209458b and HD 189733b \citep{Charbonneau2002, 2011MNRAS.416.1443S, Pont2013, 2013ApJ...774...95D}. Importantly, with a mass of only 0.19M$_{\mathrm{J}}$, WASP-127b is the most observationially accessible low-mass, gas-giant exoplanet.  Evidence of sodium, lithium, and potassium at super-solar abundances has been reported by \cite{2017A&A...602L..15P}, \cite{2018A&A...616A.145C}, and \cite{2019AJ....158..120Z}, using the ground-based NOT, GTC, and HARPS telescopes, respectively.  \cite{2017A&A...602L..15P} also report an intriguingly sharp rise in WASP-127b's transmission spectrum shortwards of 0.4$\mu$m, measured with NOT, which they attribute to a mystery UV absorber.  While this paper was under review, \cite{2020arXiv200509615S} reported an independent analysis of the same HST near-infrared spectroscopic data presented here, and measured a super-solar water abundance in WASP-127b's atmosphere. 

In this work, we present new transit observations of WASP-127b from the Hubble Space Telescope and Spitzer Space Telescope. We carried out the joint HST and Spitzer programme to observe a broad optical-to-infrared transmission spectrum for WASP-127b. The combined wavelength coverage of the programme from 0.3 to 5$\mu$m covers strong expected molecular absorption features from water and carbon-bearing species in the infrared, along with sodium and potassium absorption features, and Rayleigh scattering caused by high-altitude aerosols and H$_{2}$ in the optical region.  With this study, we aimed to classify WASP-127b as cloudy or cloud-free, and measure the abundances of important gaseous species such as H$_{2}$O, Na and K.  Our analysis confirms the super-solar abundances for WASP-127b, and we additionally find: new evidence of absorption from carbon-bearing species; and strong evidence that sub-micron sized particles are responsible for wavelength-dependent opacity from 0.3 - 1.6$\mu$m (instead of a wavelength-independent absorber).  
In the future, WASP-127b will likely become a focus of intensive James Webb Space Telescope (JWST) observations. The characterization described here will allow the community to optimize scientific objectives, instrument setup, and phase coverage for these future JWST observations.

We describe the observations and data reductions in Section 2, discuss the transit light curve fitting in Section 3, present atmospheric retrievals in Section 4, and conclude with the results and discussion in Section 5.

\section{Observations and data reduction}
The HST and Spitzer Space Telescope observations were made as part of a joint HST/Spitzer programme GO:14619 (PI: Spake).  We observed five transits of WASP-127b using different instrument setups with HST and Spitzer, in order to build a transmission spectrum covering the 0.3--5\,$\mu$m wavelength range.  In addition, the Transiting Exoplanet Survey Satellite (TESS) observed 4 photometric transits of WASP-127b in Sector 9. A summary of the observations is given in Table \ref{tab:obs}. 

\begin{table}
	\centering
	\caption{Summary of transit observations of WASP-127b.}
	\label{tab:obs}
	\begin{tabular}{lccr} % four columns, alignment for each
		\hline
		\thead{Instrument} & \thead{Start date \\ (UTC)} & \thead{Wavelength \\ range (\AA)} & \thead{Duration \\(hours)} \\
		\hline
		HST/STIS+G430L & 2018-06-23 & 2\,900$-$5\,700 & 6.8\\
		HST/STIS+G750L & 2018-02-18 & 5\,240$-$10\,270 & 6.8\\
		HST/WFC3+G141 & 2018-04-09 & 11\,000$-$17\,000 & 6.8\\
		Spitzer/IRAC Ch1 & 2017-04-02 & 31\,750$-$39\,250 & 9 \\
		Spitzer/IRAC Ch2 & 2017-04-06 & 39\,850$-$50\,050 & 9 \\
		TESS & 2019-03-05 &  6\,000$-$10\,000 & 9 \\
		TESS & 2019-03-09 &  6\,000$-$10\,000 & 9 \\
		TESS & 2019-03-18 &  6\,000$-$10\,000 & 9 \\
		TESS & 2019-03-22 &  6\,000$-$10\,000 & 9 \\
		\hline
	\end{tabular}
\end{table}

\subsection{TESS}
TESS \citep{2015JATIS...1a4003R} provides time-series photometry in a bandpass covering 0.6--1.0\,$\mu$m, and it observed four transits of WASP-127b in March 2019 (Sector 9) at 2-minute cadence.  The high cadence and multiple, opportunistic transit observations meant we could use the TESS data to refine the transit ephemeris and physical parameters of the WASP-127 system.  This was particularly useful since our HST observations do not cover much of the ingress or egress of WASP-127b's transit.  We used the TESS Presearch Data Conditioning (PDC) lightcurve of WASP-127b (Figure \ref{fig:TESS_full}), which has been corrected for effects such as non-astrophysical variability and crowding \citep{2016SPIE.9913E..3EJ}. We removed all photometric points which were flagged with anomalies, and converted the Barycentric TESS Julian Dates ($BTJD$) to $BJD_{TDB}$ by adding 2\,457\,000 days. For each of the four transits, we extracted the data in a 0.5 day window centered around the mid-transit time, and fit each transit event individually. 

\subsection{Spitzer/IRAC}
\label{sec:spitzphot}
We observed WASP-127b during two primary transits using the sub-array mode with Spitzer/IRAC channels 1 and 2, using 2 second integration times, for 9 hours each visit, with the duration set to include the 3.5 hour transit and a baseline equally as long to precisely measure the transit depth (plus some extra time as insurance). WASP-127's expected flux is 75 mJy/52 mJy for 3.6/4.5 $\mu$m, and our 2-second exposure time was short enough to stay well below saturation. The sub-array mode allowed for high cadence observations which aids in removing the detector intra-pixel sensitivity, and reduces data storage overheads. Each visit could only be done at a single wavelength requiring two transits to observe at 3.6 and 4.5 $\mu$m, as cycling between the two channels greatly exacerbates the intra-pixel sensitivity noise. Each observation began with a recommended Pointing Calibration and Reference Sensor peak-up mode of 30 minutes, which locates the star into the sub-array pixel ``sweet spot'' and helps mitigate the intra-pixel sensitivity effects providing $<$100 parts per million (ppm) accuracies. 

For both Spitzer channels, we followed the data reduction and photometry procedures of \cite{2015MNRAS.451..680E}.  We reduced the Basic Calibrated Data (BCD) frames for each light curve using a publicly-available PYTHON pipeline\footnote{from www.github.com/tomevans}, which does the following: first, it calculates the background level and locates the stellar centroid in each BCD frame. It estimates the background from the median pixel value of four 8\,$\times$ \,8 pixel subarrays at the corners of each frame, and then subtracts that value from each pixel in the array. It finds the centroid coordinates by taking the flux-weighted mean of a 7\,$\times$\,7 pixel subarray centred on the star.  The pipeline computes exposure mid-times in Barycentric Julian Date Coordinated Universal Time (BJD$_{UTC}$) using the $BMJD_{OBS}$ and $FRAMTIME$ header entries. It flags bad frames by identifying frames whose centroid coordinates or pixel counts deviate by 5$\sigma$ from those of the 30 frames immediately preceding and following each frame. We removed bad frames from the analysis. We iterated this bad-frame identification twice, and discarded less than 5\,\% of the frames.

The pipeline performed photometry on each remaining frame by summing the pixel counts within circular apertures of various sizes between 1.5 and 6 pixels, in increments of 0.5 pixels.  Because the IRAC point spread function (PSF) is undersampled, we linearly interpolated the pixel array on to a 10\,$\times$\,10 supersampled grid, which has previously been done by \cite{2010Natur.464.1161S}, for example. We counted the interpolated subpixels towards the aperture sum if their centres fell within the aperture radius.  Our selected photometric light curves are shown in Figures \ref{fig:w127_spitzlc1} and \ref{fig:w127_spitzlc2}, and we discuss how the apertures were chosen in Section \ref{spitzfit}.

\subsection{STIS}
We observed two transits with HST's Space Telescope Imaging Spectrograph (STIS), one each with the G430L and G750L gratings. We followed an observing strategy proven to produce high signal-to-noise spectra (e.g. \citealt{2001ApJ...552..699B}, \citealt{2011MNRAS.416.1443S}, \citealt{2012MNRAS.422.2477H}, \citealt{2015MNRAS.447..463N}). The data were taken on 2018-06-23 and 2018-02-18, covering wavelengths of 2\,900--5\,700 and 5\,240--10\,270\AA, respectively.  Visits 1 and 2 both lasted 4.5 spacecraft orbits each.  One HST orbit lasts $\sim$\,96 minutes during which WASP-127b is visible for $\sim$\,45 minutes, leaving $\sim$\,45 minute gaps in the data as the spacecraft passes through the Earth's shadow.  WASP-127b has a long transit duration ($\sim$\,3.5 hours, compared to $\sim$\,2 hours for a typical hot Jupiter, e.g. HD\,209458b).  We scheduled each visit such that 2 orbits fell fully inside a transit and 1.5 fell either side of it, in order to accurately measure the baseline stellar flux.  We used integration times of 280 and 180 seconds, resulting in a total of 48 and 58 low-resolution spectra ($\Delta \lambda / \lambda = 500$) for the G430L and G750L visits respectively.  We used 52"\,x\,2" slits to minimise slit losses, and minimised the data-acquisition overheads by reading out a smaller portion of the CCD (128\,x\,128 pixels).

Our data reduction method for STIS follows previous works such as \cite{2013MNRAS.436.2956S}, \cite{2013MNRAS.434.3252H}, and \cite{2014MNRAS.437...46N, 2015MNRAS.447..463N}.  We used the most recent version of the CALSTIS automatic reduction pipeline \citep{1998stis.rept...14K} included in IRAF\footnote{IRAF is distributed by the National Optical Astronomy Observatories, which are operated by the Association of Universities for Research in Astronomy, Inc., under cooperative agreement with the National Science Foundation.} \citep{1993ASPC...52..173T} to reduce the raw STIS data (which involves bias-, dark-, and flat-correction). Similarly to \cite{2015MNRAS.447..463N}, we corrected the G750L spectra for fringing effects with the method described in \cite{1998stis.rept...29G}.  Further, we used the method described in \cite{2013A&A...553A..26N} to correct the data for cosmic rays and bad pixels flagged by the CALSTIS pipeline.  We then extracted 1D spectra from the reduced data frames using IRAF's APALL.  We used aperture widths ranging from 3.5 to 10.5 pixels, in 1-pixel steps, and found the aperture width for each visit that gave the lowest residual scatter in the white light curve (see Section \ref{sec:w127specfits}).  The selected aperture widths were 9.5 and 10.5 pixels for G430L and G750L, respectively.  Finally, the wavelength solutions for each spectrum were obtained from the $x1d$ files from CALSTIS, and the spectra were then cross-correlated with the median of the out-of-transit spectra to place them on a common wavelength scale, which helps to account for sub-pixel shifts in the dispersal direction.

\subsection{WFC3}
We observed one spectroscopic transit of WASP-127b using HST's Wide Field Camera 3 (WFC3) instrument with the infrared G141 grism.  The observations spanned the approximate wavelength range of 11\,000--17\,000\AA$\,$, which covered a broad band of water absorption lines centred on 14\,000\AA$\,$.  We used HST's spatial scan mode and a scan rate of 1 pixel per second for 15 observations of 120 seconds each, which spread WASP-127's spectrum over 120 pixels. We used the SPARS10 sampling sequence with 14 non-destructive reads per exposure (NSAMP = 14).  The maximum number of electron counts per pixel was 29\,000 - which is approximately 40\% of the saturation limit of the detector.    
%We note that the HST observations proved difficult to schedule in practice, due to two factors.  Firstly, observations cannot be made while the spacecraft is crossing the South Atlantic Anomaly (SAA).  Because the transit duration of WASP-127b is long (approximately 3.5 hours) 5 telescope orbits were required to cover the transit and baseline either side.  It was difficult to find 5 consecutive orbits centred on a transit event that fell between SAA crossings, particularly as WASP-127b transits only once every 4.2 days.  Secondly, WASP-127 has two nearby stars (XX, X"; ZZ, Z") whose spectra need to be arranged on the detector such that they do not overlap with that of WASP-127.  We therefore placed strict roll-angle constraints on the telescope, which further reduced the observation opportunities.  To mitigate these difficulties, half an orbit of each HST observation was sacrificed to provide increased scheduling flexibility. 
The raw frames were first reduced with the automatic CalWF3 pipeline. The 1-D spectra were then extracted following standard methods (e.g \citealp{2017Natur.548...58E}), i.e. by building up flux counts by summing the difference between successive non-destructive reads.  First, we removed the background from each read difference by subtracting the median of a box of pixels uncontaminated by the spectrum.  Then, we found the flux-weighted centre of each read-difference and set to zero all pixels more than 70 rows away from the centre in the cross-dispersion axis, which removes a majority of the cosmic rays. The remaining cosmic rays were flagged by finding 4$\sigma$ outliers relative to the median along the dispersion direction. We replaced each flagged pixel with the median along the dispersion direction, re-scaled to the count rate of the cross-dispersion column.  We then summed the corrected read differences to produce background- and cosmic-ray-corrected 2D images of the spatially scanned spectrum.  We extracted 1D spectra from the corrected images by summing the flux in a rectangular aperture, centred on the flux-weighted centre of the scan.  The aperture spanned the full dispersion axis and was 140 pixels wide in the spatial direction.

We found the wavelength solutions by cross-correlating the extracted spectra with an ATLAS model stellar spectrum \citep{2004astro.ph..5087C} which most closely matches WASP-127 ($T_{\mathrm{eff}}$ = 5\,500 K, log $g$ = 4.0 cgs); modulated by the G141 grism throughput. Following standard methods \citep{2018ApJ...858L...6K} we interpolated each spectrum onto the wavelength range of the first to account for shifts in the dispersion axis over time.

\section{Light curve fitting}
\subsection{TESS}
We fit each of the four TESS transit light curves separately, using the 4-parameter, non-linear, limb-darkened transit model of \cite{MandelAgol2002}, multiplied by a linear baseline time trend. We used the method of \cite{2010A&A...510A..21S} to estimate the limb-darkening coefficients, using the ATLAS model stellar spectrum \citep{2004astro.ph..5087C} which most closely matched WASP-127b ($T_{\mathrm{eff}}$ = 5\,500 K, log $g$ = 4.0 cgs).  We found limb-darkening coefficients of $c_{1}$ = 0.5802, $c_{2}$ = -0.1496, $c_{3}$ = 0.5504, and $c_{4}$ = -0.3115. 

For each of the 4 transits, we fit for six free parameters: the central transit time; planet-to-star radius ratio ($R_{P}/R_{\mathrm{*}}$); two coefficients for the linear baseline; the cosine of the orbital inclination ($i$); and $a/R_{\mathrm{*}}$ (where $a$ is the semi-major axis and $R_{\mathrm{*}}$ the stellar radius). We found weighted-average parameter values, including the radius ratio of $R_{P}/R_{\mathrm{*}}$ = 0.10060 $\pm$ 0.00038; orbital inclination of $i$=87.6$\pm$1.0 degrees; and $a/R_{\mathrm{*}}$=7.83$\pm$0.30. The inclination and $a/R_{\mathrm{*}}$ were used as fixed values in the HST and Spitzer analyses. We found the four transit mid-times to be 2458548.120564 $\pm$ 0.000357, 2458552.297877 $\pm$ 0.000349, 2458560.654727 $\pm$ 0.000351, and 2458564.832716 $\pm$ 0.000342, in $BDJ_{TBD}$.  The four individually-measured radius ratios were $0.10105 \pm 0.00109$, $0.10088 \pm 0.00105$, $0.10067 \pm 0.00111$, and $0.10043 \pm 0.00049$.  The lack of variation in both the radius ratio and the observed stellar flux over the TESS observational period (Figures \ref{fig:TESS_full} and \ref{fig:TESS_tdepths}) increases the evidence that WASP-127 is a photometrically quiet star.  Therefore, we should not expect stellar activity to vary the transit depths measured at different epochs.

\begin{figure}
    \centering
    \includegraphics[width=0.49\textwidth]{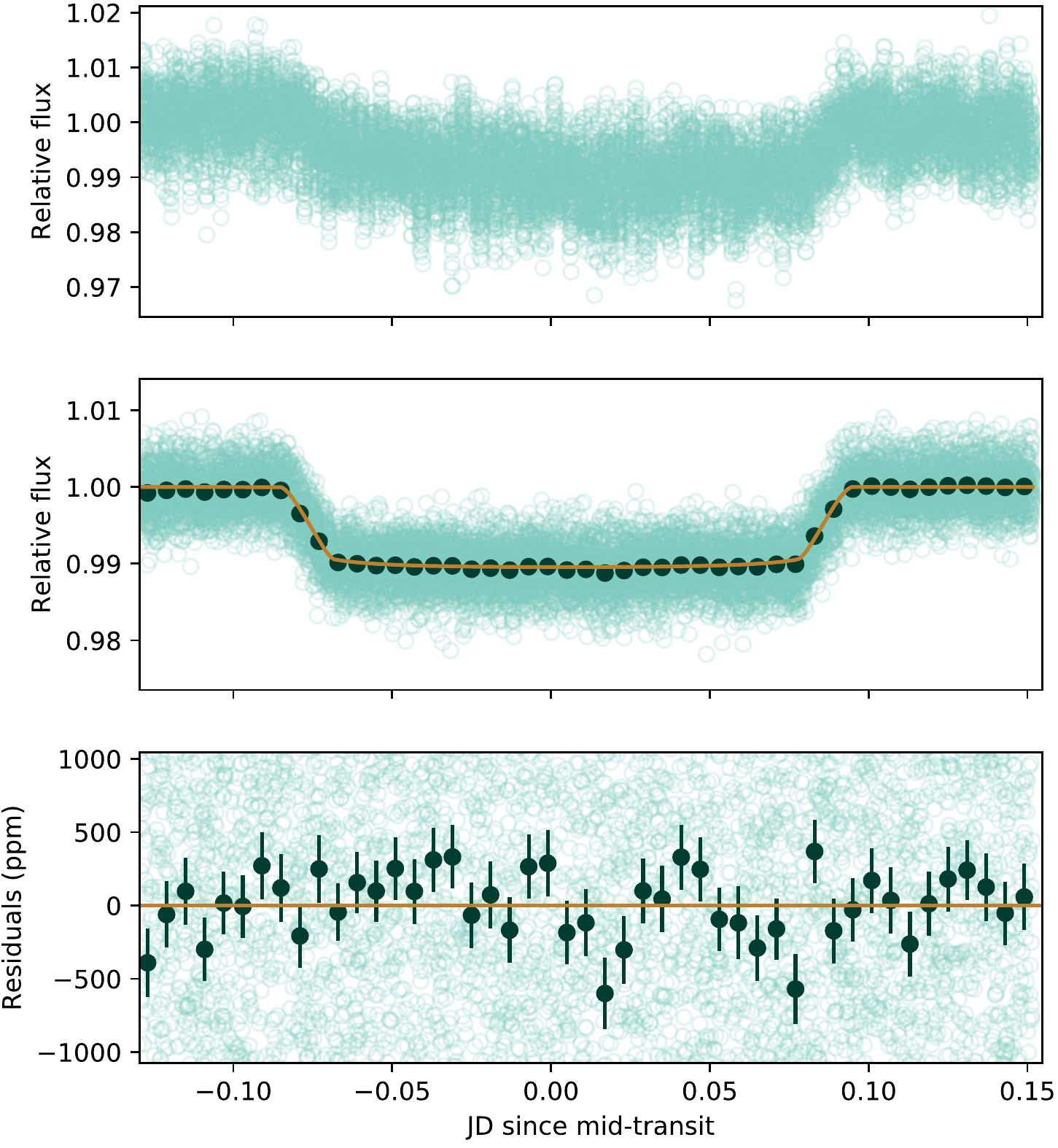}
    \vspace*{-1\baselineskip}          %Use command to remove white space
    \caption{Light curve fit for WASP-127b using Spitzer/IRAC's 3.6$\mu$m channel. Top panel: light green points are raw data.  Middle panel: light green points are data divided by systematics model, dark green points are data in 9-minute bins for clarity, beige curve is the best-fit transit model. Bottom panel: Best-fit model residuals.}
    \label{fig:w127_spitzlc1}
\end{figure}

\begin{figure}
    \centering
    \includegraphics[width=0.49\textwidth]{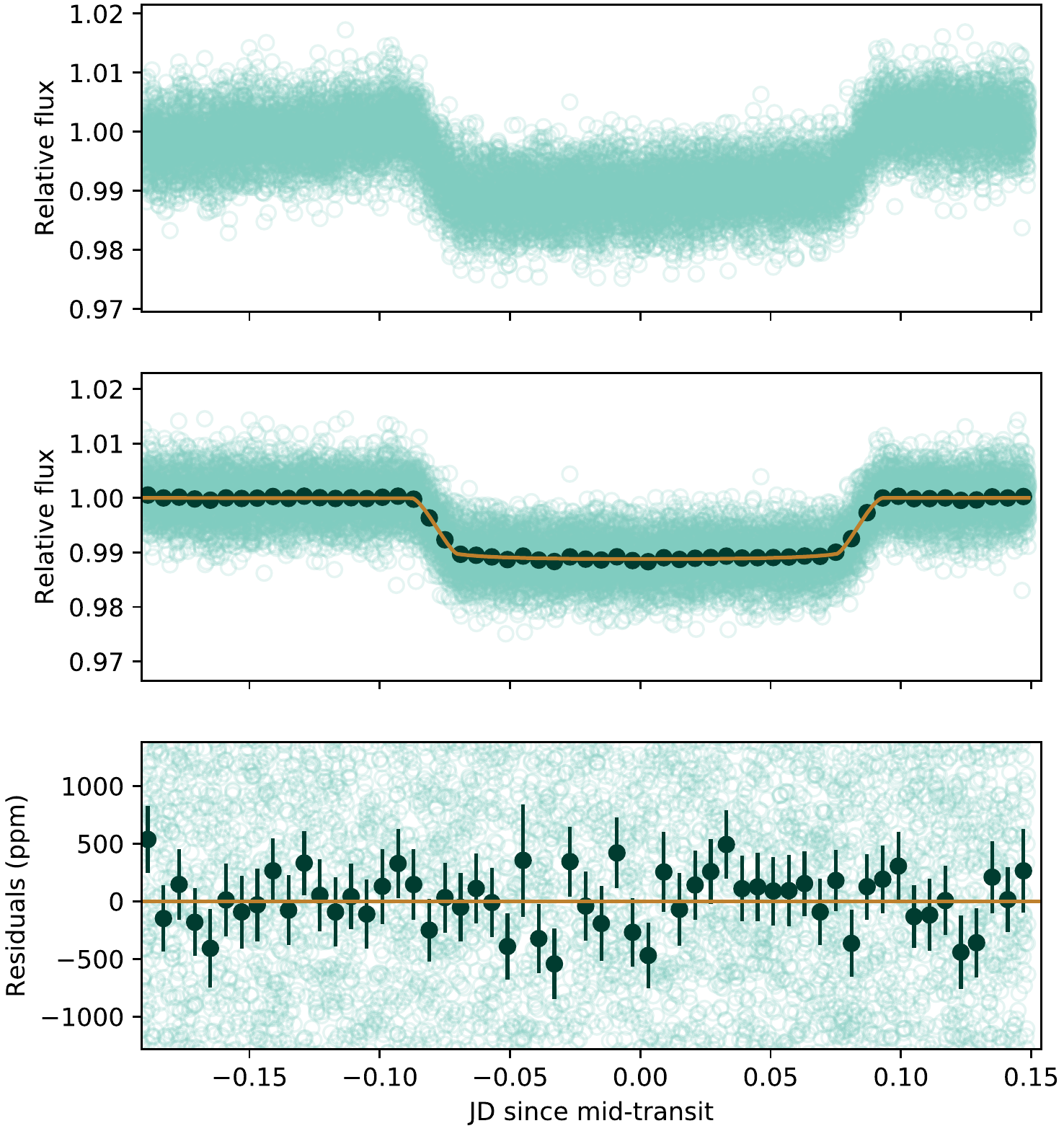}
    \vspace*{-1\baselineskip}          %Use command to remove white space
    \caption{Light curve fit for WASP-127b using Spitzer/IRAC's 4.5$\mu$m channel. Top panel: light green points are raw data.  Middle panel: light green points are data divided by systematics model, dark green points are data in 9-minute bins for clarity, beige curve is the best-fit transit model. Bottom panel: Best-fit model residuals.}
    \label{fig:w127_spitzlc2}
\end{figure}

\subsection{Spitzer}
\label{spitzfit}
We used the same light curve fitting procedure for both Spitzer/IRAC channels. Spitzer photometry is prone to large intra-pixel systematics (e.g. \citealt{2005Natur.434..740D}).  To correct for this we fit for a two-dimensional quadratic trend in the photometry with the $x$ and $y$ position of WASP-127's centroid (measured using the pipeline discussed in Section \ref{sec:spitzphot}).  The function has the form 
\begin{equation}
    F = c_{\mathrm{2,x}} x^{2} + c_{\mathrm{2,y}} y^{2} + c_{\mathrm{1,x}} x + c_{\mathrm{1,y}} y +  c_{\mathrm{xy}} xy,
\end{equation}
\noindent and we fit for the following five free parameters: $c_{\mathrm{2,x}}$, $c_{\mathrm{1,x}}$, $c_{\mathrm{2,y}}$, $c_{\mathrm{1,y}}$ and $c_{\mathrm{x,y}}$.
We used the BATMAN Python package \citep{2015PASP..127.1161K} to model the transit light curve signal, and fit for the planet-to-star radius ratio ($R_{\mathrm{p}}/R_{\mathrm{*}}$) and mid-transit time ($t_{\mathrm{0}}$).  We also fit for the gradient ($c_{\mathrm{1}}$) and linear trend ($c_{\mathrm{0}}$) in the baseline in the photometry.  In total there were 9 free parameters in the light curve fit.  We used the Markov chain Monte Carlo (MCMC) package emcee \citep{2013PASP..125..306F} to marginalise over the parameter space of the model likelihood distribution. We used 100 walkers and ran chains for 5\,000 steps, discarding the first 1\,000 as burn-in before combining the walker chains into a single chain.  We followed this procedure for each of the photometric lightcurves that we produced, which used varying aperture sizes from a radius of 1.5 to 6 pixels, in increments of 0.5 pixels.  Here we quote the results from the lightcurve which had the lowest model residuals after the fitting process.  For the 3.6 $\mu$m channel the optimum aperuture radius was 3.0 pixels, and for the 4.5 $\mu$m channel it was 2.5 pixels.  Table \ref{tab:tab:spitz} shows our best-fit transit depths and mid-transit times for each channel.  Figures \ref{fig:w127_spitzlc1} and \ref{fig:w127_spitzlc2} show the light curves with their best-fit models and residuals, and Figures \ref{fig:w127_ch1triangle} and \ref{fig:w127_ch1triangle} show the posterior distributions of these fits.

\begin{table*}
\centering
\begin{tabular}{lcc}
\hline
Parameter & Value\\
\hline
Period (day) & 4.178$^{a}$ \\
$a/R_{\mathrm{*}}$ & 8.044$^{a}$ \\
Inclination (\degree) & 88.7$^{b}$ \\
Eccentricity & 0$^{b}$ \\
Arg. of Periastron & 90$^{b}$ \\
\hline
 & Channel 1  & Channel 2\\
\hline
Transit depth (\%) & 0.993$^{+0.005}_{-0.005}$ & 1.073$^{+0.006}_{-0.006}$ \\
Mid-time (JD) & 2\,457\,846.19996$^{+0.00004}_{-0.00003}$ & 2\,457\,850.37968$^{+0.00001}_{-0.00001}$  \\
LD coefficients ($u_{1,2}$) & 0.0626$^{c}$, 0.1734$^{c}$ & 0.0639$^{c}$, 0.1374$^{c}$ \\
\hline
\end{tabular}
\caption{Results from light curve fit for WASP-127b using Spitzer/IRAC Channel 1 (3.6$\mu$m) and Channel 2 (4.5$\mu$m).  $^{a}$ period and transit mid-times fixed to values found from our updated fit described in Section \ref{sec:ephemfit}.  $^{b}$ planet parameters fixed to values from \citep{2017A&A...599A...3L}.  $^{c}$ Limb darkening parameters fixed from ATLAS models \citep{2004astro.ph..5087C}.}
\label{tab:spitz}
\end{table*}

Our results were consistent with those inferred by a more sophisticated treatment of Spitzer's systematics.  \cite{2015ApJ...805..132D} use a technique that involves modelling the light curves of individual pixels to correct for Spitzer's intra-pixel variations, called Pixel Level Decorrelation (PLD).  The measured transit depths using both PLD and the light-curve fitting procedure described above are shown in Figure \ref{fig:w127_spec}.  For both channels they are consistent within 1$\sigma$.

\subsection{Period and ephemeris fitting}
\label{sec:ephemfit}
\begin{figure}
    \centering
    \includegraphics[width=\columnwidth]{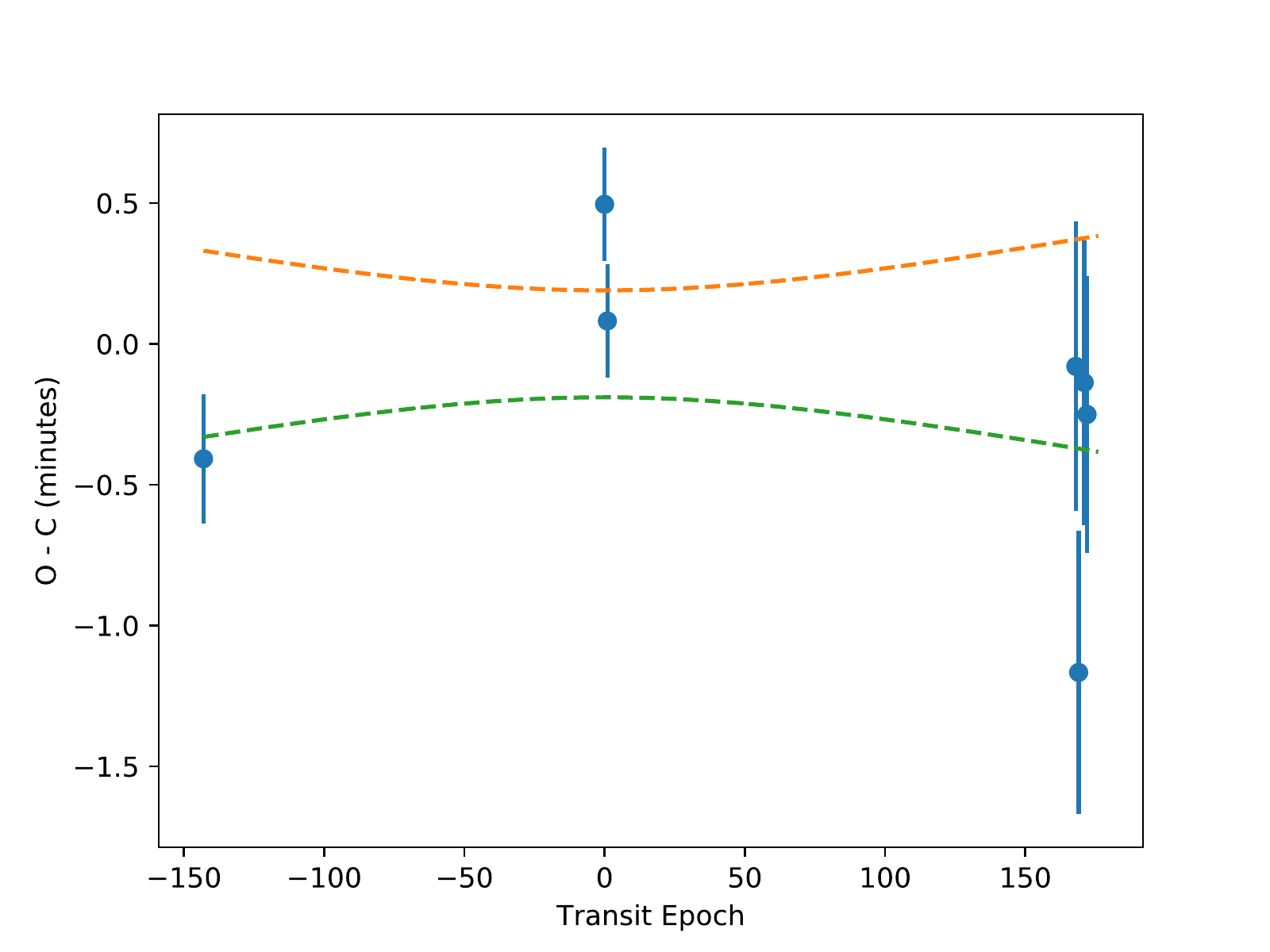}
    \vspace*{-1\baselineskip}          %Use command to remove white space
    \caption{Timing offsets from the fitted mid-transit times for
    WASP-127b.  Label discovery paper, TESS and Spitzer data.  Transit epoch centred on Spitzer data. }
    \label{fig:w127_timing}
\end{figure}

WASP-127b has a long, $\sim$3 hour transit, which, combined with the Earth occultations occurring in each HST orbit, meant we are unable to get continuous phase coverage of the target over the full transit.  Because of the particular timings of our observations, we did not observe much of the ingress or the egress on any of the three HST transit observations.  This made it difficult to fit for the transit mid-time.  Indeed, when we did fit for the transit mid-time our best-fit solution for the WFC3 visit was earlier than expected by 6 minutes.  Large inaccuracies in the mid-transit time can can change the measured transit depth.  However, the TESS and Spitzer observations have high cadence and full phase coverage, and so their mid-transit times may be more reliable.  They can also be used to update the discovery-paper period and ephemeris so that the mid-transit times for the HST visits can be fixed to more reliable values.  We fit the discovery paper reported values, all four TESS transit times and two spitzer transit times with a linear trend in time, fitting for the period and ephemeris (Figure \ref{fig:w127_timing}).  Our best fit period was 4.17807 $\pm$ 0.00013 days, and our best-fit ephemeris was 2457846.20526 $\pm$ 0.00013.  The updated period and ephemeris were used to fix the mid-transit times for all three HST transit observations.

\begin{figure}
    \centering
    \includegraphics[width=0.49\textwidth]{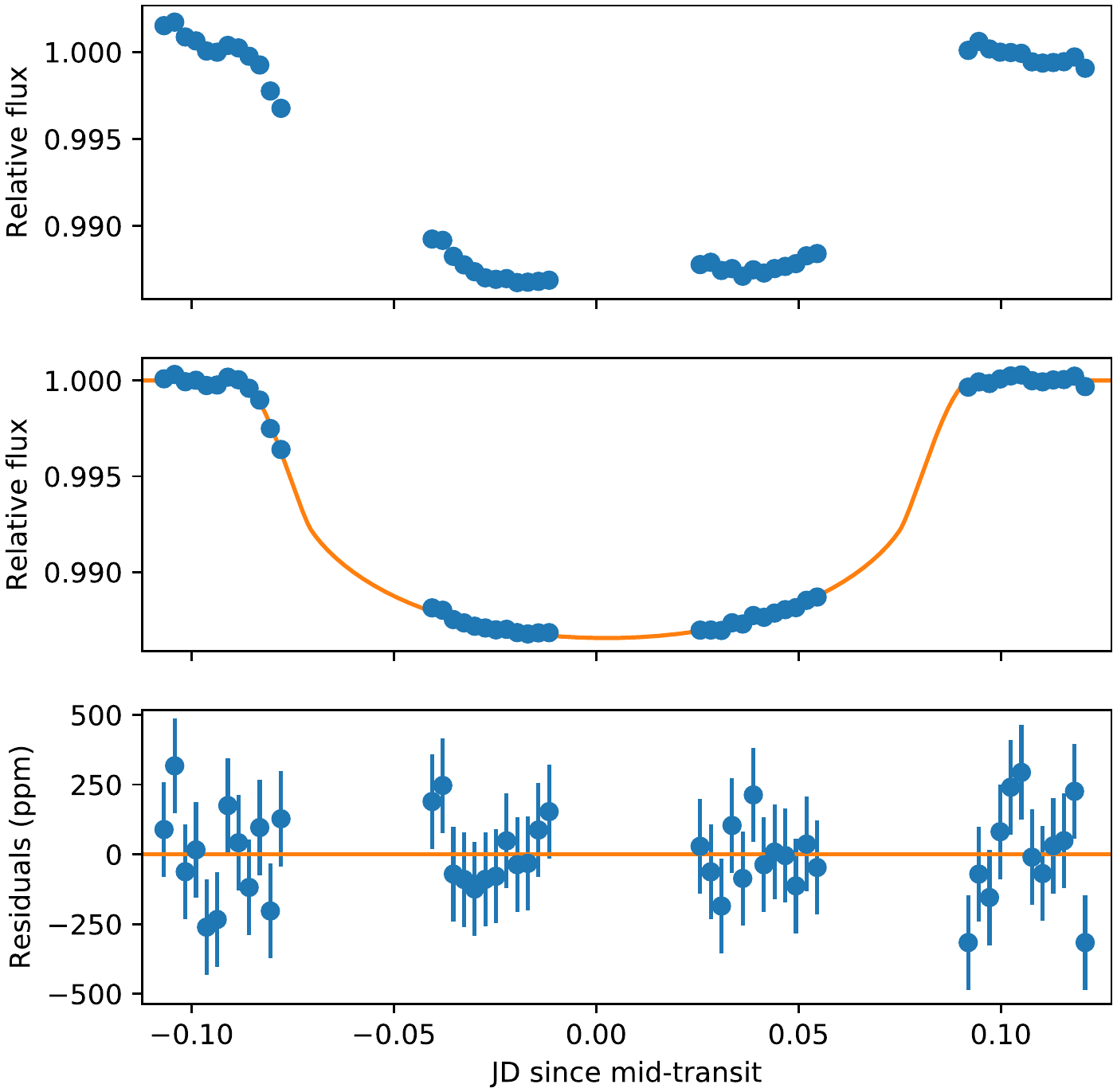}
    \vspace*{-1\baselineskip}          %Use command to remove white space
    \caption{White light curve fit for WASP-127b using HST/STIS+G430L, covering the entire 2\,900$-$5\,700\AA$\,$ wavelength range. Top panel: raw flux before de-trending, divided by the median of the out-of-transit data.  Middle panel: points are data divided by systematics model, curve is the best-fit transit model. Bottom panel: best-fit model residuals.}
    \label{fig:w127_g430white}
\end{figure}

\subsection{HST: STIS and WFC3}
We followed the same light-curve fitting procedure for all three of the STIS and WFC3 visits, but here describe the process for one visit.  
\subsubsection{White light curve fit}
\begin{figure}
    \centering
    \includegraphics[width=0.49\textwidth]{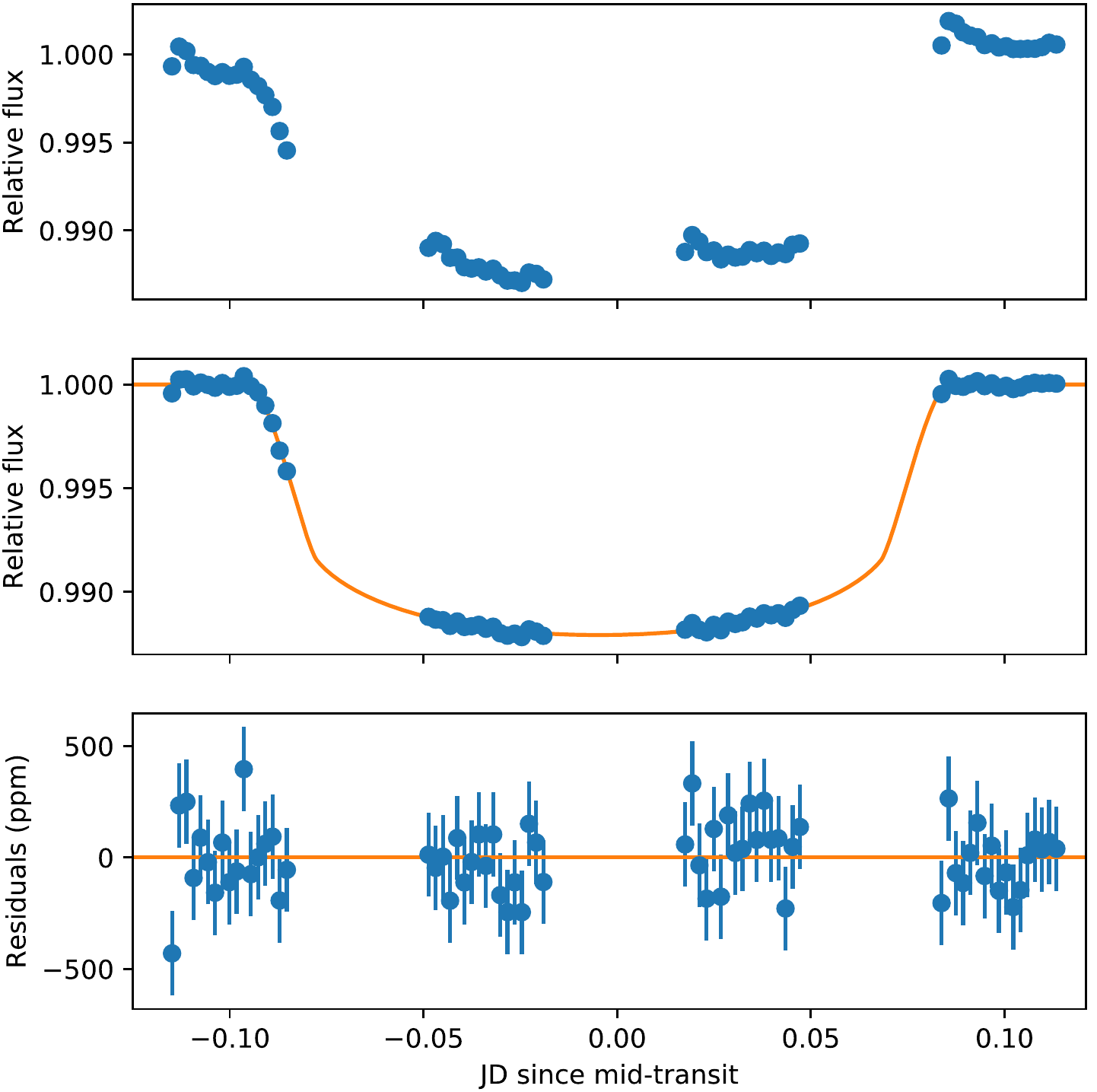}
    \vspace*{-1\baselineskip}          %Use command to remove white space
    \caption{White light curve fit for WASP-127b using HST/STIS+G750L, covering the entire 5\,240$-$10\,270\AA$\,$ wavelength range. Top panel: raw flux before de-trending, divided by the median of the out-of-transit data.  Middle panel: points are data divided by systematics model, curve is the best-fit transit model. Bottom panel: best-fit model residuals.}
    \label{fig:w127_g750white}
\end{figure}

\begin{figure}
    \centering
    \includegraphics[width=0.49\textwidth]{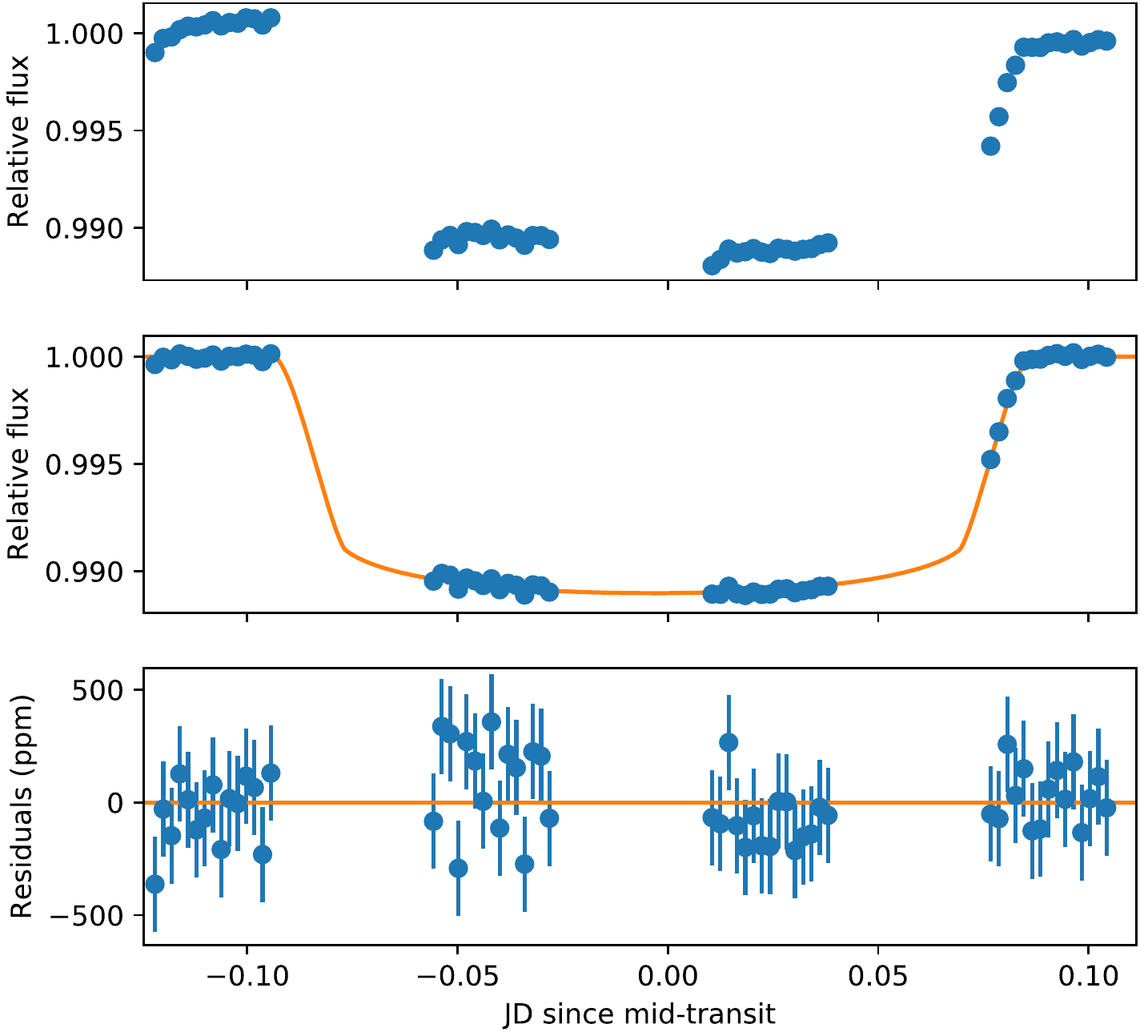}
    \vspace*{-1\baselineskip}          %Use command to remove white space
    \caption{White light curve fit for WASP-127b using HST/WFC3+G141, covering the entire 11\,000 - 17\,000\AA$\,$ wavelength range. Top panel: raw flux before de-trending, divided by the median of the out-of-transit data.  Middle panel: points are data divided by systematics model, curve is the best-fit transit model. Bottom panel: best-fit model residuals.}
    \label{fig:w127_wfc3white}
\end{figure}

For each visit (which comprises a set of time-series 1D spectra), we first created a white light curve by summing the counts of the 1D spectra across all wavelengths.  The resulting time-series flux measurements show the transit signal modulated by systematic trends which correlate with HST phase, and the changing position of the spectrum on the detector.  Such trends are commonly reported in STIS and WFC3 time-series data (e.g. \citealt{2001ApJ...552..699B}, \citealt{2013ApJ...774...95D,2016ApJ...819...10W}).  Since we do not know the functional form of the systematic trends, \cite{2012MNRAS.419.2683G} suggest treating the lightcurve as a Gaussian Process (GP).  Therefore, we follow the implementation of GPs for the HST lightcurves pioneered by \cite{2013ApJ...772L..16E, 2018AJ....156..283E}, except we use the Python library for GP regression, George \citep{hodlr} rather than a custom code.  Similarly to \cite{2013ApJ...772L..16E, 2018AJ....156..283E}, we used a squared- exponential kernel for the GP covariance matrix.  We used three GP input variables - the HST orbital phase ($\phi$), the position of the spectrum in the spatial direction on the detector ($x$), and the position in the dispersion direction ($y$).  This gave four free GP parameters: the covariance amplitude ($A$), and a correlation length scale for each of the four input variables: $L_{\mathrm{\phi}}$, $L_{\mathrm{x}}$, and $L_{\mathrm{y}}$ for HST phase, $x$, and $y$ respectively.  We used BATMAN to model the transit light curve signal, and fit for the planet-to-star radius ratio ($R_{\mathrm{p}}/R_{\mathrm{*}}$), fixing the remaining orbital parameters to the values given in Table \ref{tab:wlg7}. To model the stellar limb darkening we fitted a four-parameter non-linear limb darkening law \citep{2000A&A...363.1081C} to the ATLAS stellar model  \citep{2004astro.ph..5087C} which most closely matches WASP-127 ($T_{\mathrm{eff}}$ = 5\,500 K, log $g$ = 4.0 cgs).  We also fit for the gradient ($c_{\mathrm{1}}$) and y-intercept ($c_{\mathrm{0}}$) of a linear trend in the out-of-transit baseline.  Therefore, for the white light curve, we fit for 7 free parameters overall: $R_{\mathrm{p}}/R_{\mathrm{*}}$, $c_{\mathrm{1}}$, $c_{\mathrm{0}}$, $A$, $L_{\mathrm{\phi}}$, $L_{\mathrm{x}}$, and $L_{\mathrm{y}}$. We used the Markov chain Monte Carlo (MCMC) package emcee \citep{2013PASP..125..306F} to marginalise over the parameter space of the model likelihood distribution. We used 80 walkers and ran chains for 500 steps, discarding the first 100 as burn-in before combining the walker chains into a single chain. The best-fit results for the transit depths $([R_{\mathrm{p}}/R_{\mathrm{*}}]^{2})$ are given in Tables \ref{tab:wlg4}, \ref{tab:wlg7} and \ref{tab:whwf} for the G430L, G750L and G141 visits respectively.  Similarly, Figures \ref{fig:w127_g430white}, \ref{fig:w127_g750white} and \ref{fig:w127_wfc3white} show the best-fit white light curves and their residuals for each visit.  Figures \ref{fig:w127_g430triangle_white}, \ref{fig:w127_g750triangle_white}, and \ref{fig:w127_wfc3triangle_white} show corner plots of the MCMC chains, which illustrate the posterior distributions for each of the fits.  The posterior distributions appear well sampled, and there are no problematic correlations between $R_{\mathrm{p}}/R_{\mathrm{*}}$ and the other fitted parameters.

\begin{table*}
    \centering
    \begin{tabular}{lccc}
    \hline
    Parameter & & Value \\
    \hline
    $a/R_{\mathrm{*}}$ & & 8.044$^{a}$ \\
    Inclination (\degree) & & 88.7$^{a}$ \\
    Eccentricity & & 0$^{a}$ \\
    Arg. of Periastron & & 90$^{a}$ \\
    Period (day) & & 4.178$^{b}$ \\
    \hline
    & STIS+G430L & STIS+G750L & WFC3+G141 \\
    \hline 
    Transit depth (\%) & 1.034$^{+0.006}_{-0.005}$ & 1.013$^{+0.009}_{-0.006}$ & 0.996$^{+0.011}_{-0.011}$ \\
    Mid-time (JD) & 2\,458\,293.2528$^{b}$ & 2\,458\,293.2528$^{b}$ & 2\,458\,293.2528$^{b}$ \\
    $u_{1}$ & 0.5466$^{c}$ &  0.7017$^{c}$ & 0.5944$^{c}$\\
    $u_{2}$ &  -0.3781$^{c}$ & -0.5462$^{c}$ & 0.0707$^{c}$\\
    $u_{3}$ & 1.2964$^{c}$ & 1.1008$^{c}$ & -0.1204$^{c}$\\
    $u_{4}$ & -0.5955$^{c}$ & -0.5233$^{c}$ & 0.0202$^{c}$ \\
    \hline
    \end{tabular}
\caption{Fixed planet parameters and results from white light curve fits for WASP-127b using HST. $^{a}$ planet parameters fixed to values from \citep{2017A&A...599A...3L}. $^{b}$ period and transit mid-times fixed to values found from our updated fit described in Section \ref{sec:ephemfit}.  $^{c}$ Limb darkening parameters fixed from ATLAS models \citep{2004astro.ph..5087C}. }
\label{tab:wltab}
\end{table*}

\begin{figure*}
    \centering
    \includegraphics[width=0.8\textwidth]{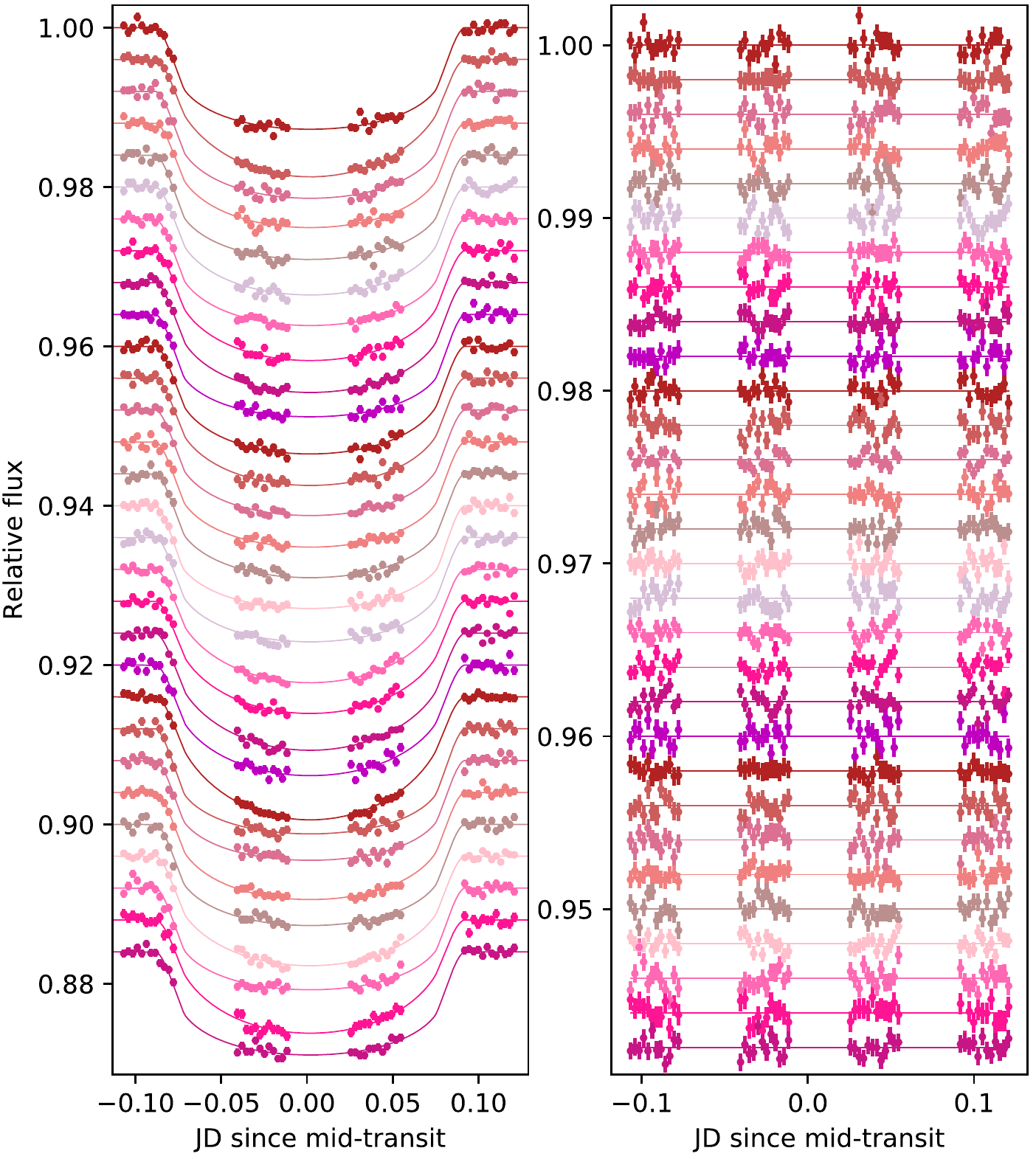}
    \vspace*{-1\baselineskip}          %Use command to remove white space
    \caption{Spectroscopic light curves for WASP-127b using HST/STIS+G430L, covering the 2\,900$-$5\,700\AA$\,$ wavelength range. (a) Points are light curves divided by systematics models, curves are best-fit transit models. (b) Best-fit model residuals.  Arbitrary vertical offsets applied for clarity.}
    \label{fig:w127_g430speclcs}
\end{figure*}

\begin{figure*}
    \centering
    \includegraphics[width=0.8\textwidth]{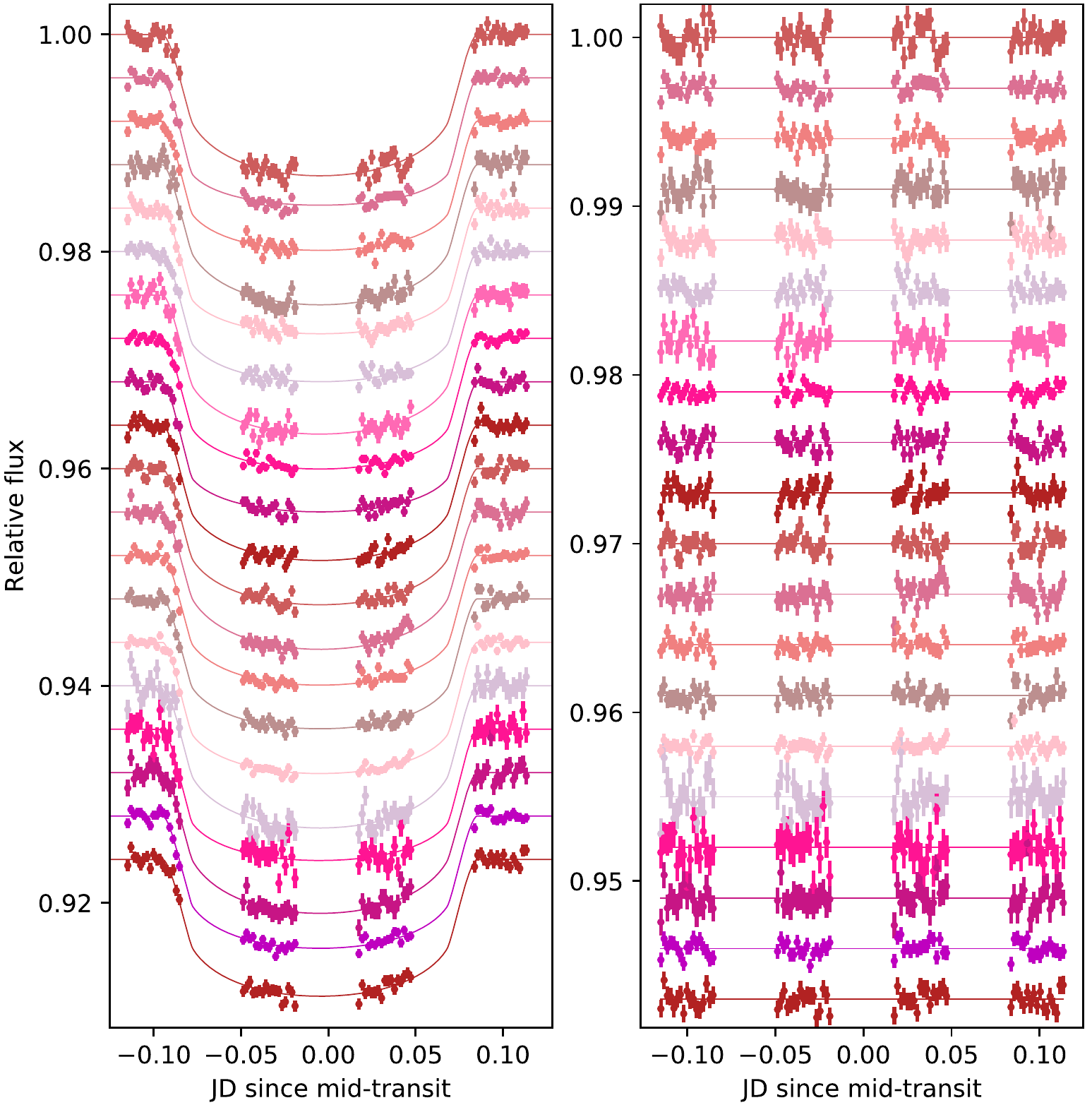}
    \vspace*{-1\baselineskip}          %Use command to remove white space
    \caption{Spectroscopic light curves for WASP-127b using HST/STIS+G750L, covering the 5\,240$-$10\,270\AA$\,$ wavelength range. (a) Points are light curves divided by systematics models, curves are best-fit transit models. (b) Best-fit model residuals.  Arbitrary vertical offsets applied for clarity.}
    \label{fig:w127_g750speclcs}
\end{figure*}

\begin{figure*}
    \centering
    \includegraphics[width=0.8\textwidth]{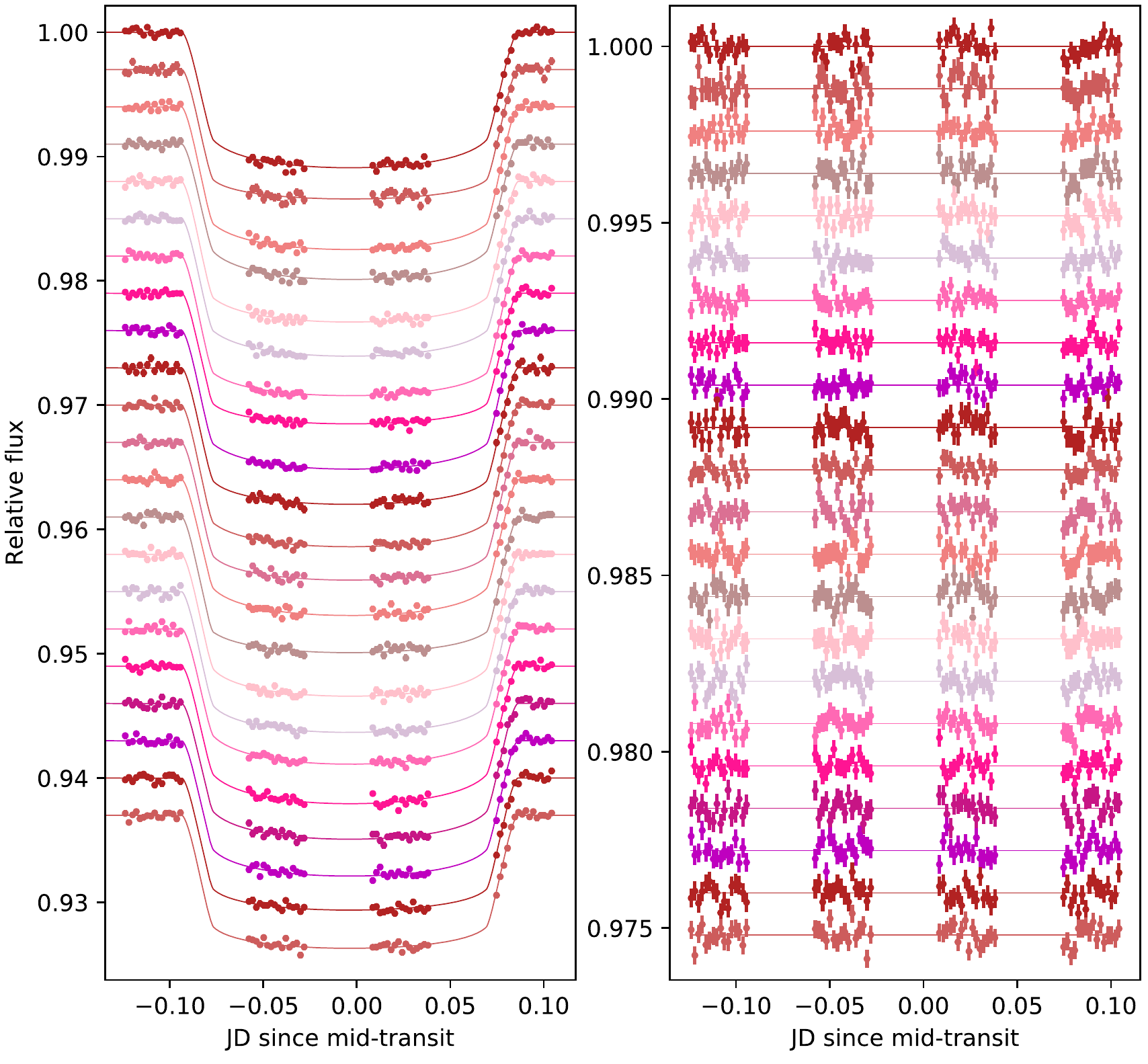}
    \vspace*{-1\baselineskip}          %Use command to remove white space
    \caption{Spectroscopic light curves for WASP-127b using HST/WFC3+G141, covering the 11\,000-17\,000\AA$\,$ wavelength range. (a) Points are light curves divided by systematics models, curves are best-fit transit models. (b) Best-fit model residuals.  Arbitrary vertical offsets applied for clarity.}
    \label{fig:w127_wfc3speclcs}
\end{figure*}

\subsubsection{Spectroscopic light curve fit}
\label{sec:w127specfits}
We used the same spectroscopic light curve fitting procedure for each of the three HST visits. First, we binned each individual spectrum into spectroscopic channels (the wavelength ranges are given in Tables \ref{tab:w127_g430l}, \ref{tab:w127_g750l} and \ref{tab:w127_wfc3}), to make the spectroscopic light curves.  We then employed a ``common-mode'' correction to remove wavelength-independent systematic trends in the data (e.g. \citealt{2018AJ....156..283E}). The process entails finding the systematic trends common to all wavelengths by dividing the white light curve by its best-fit transit model, and then subtracting the result from each spectroscopic lightcurve before the fitting procedure.  Similarly to the white light curve fits, we fixed $t_{0}$ to the expected values found from our updated period and ephemeris, and fixed the orbital parameters to those derived from \cite{2017A&A...599A...3L}, and wavelength-dependent limb darkening coefficients from the same ATLAS model which best describes WASP-127. Similarly to the white light curve fit, we treat each spectroscopic light curve as a Gaussian Process, using a squared-exponential kernel, and the same three GP input varables: ($\phi$), ($x$), and ($y$).
Therefore, for each spectroscopic channel the fitted parameters were $R_{\mathrm{p}}/R_{\mathrm{*}}$, $c_{\mathrm{1}}$, $c_{\mathrm{0}}$, $A$, $L_{\mathrm{\phi}}$, $L_{\mathrm{x}}$, and $L_{\mathrm{y}}$. We used the Markov chain Monte Carlo (MCMC) package emcee \citep{2013PASP..125..306F} to marginalise over the parameter space of the model likelihood distribution. We used 80 walkers and ran chains for 500 steps, discarding the first 100 as burn-in before combining the walker chains into a single chain.  The best-fit transit depths are given in Tables \ref{tab:w127_g430l}, \ref{tab:w127_g750l} and \ref{tab:w127_wfc3} for the G430L, G750L and G141 visits respectively.  The best fit spectroscopic lightcurves and their residuals are shown in Figures \ref{fig:w127_g430speclcs}, \ref{fig:w127_g750speclcs} and \ref{fig:w127_wfc3speclcs}.  Example posterior distributions for individual spectroscopic light curves are shown in Figures \ref{fig:w127_g430triangle}, \ref{fig:w127_g750triangle} and \ref{fig:w127_wfc3triangle}.  The posterior distributions appear well sampled, and there are no problematic correlations between $R_{\mathrm{p}}/R_{\mathrm{*}}$ and the other fitted parameters. We also ran the spectroscopic analysis for WFC3 G141 following the instrument systematic marginalization method outlined in \cite{2016ApJ...819...10W} and found the transmission spectral shape to be robust across the two systematic treatments with the absolute values of the measured depths within 1-$\sigma$ of each other.  Similarly, we ran the spectroscopic anslysis for both STIS visits following the marginalisation method outlined in \cite{2014MNRAS.437...46N}, and found that the transit depths agreed within 1-$\sigma$.

\begin{table*}
\centering
\begin{tabular}{cccccccc}
\hline
   Bin start (\AA) &   Bin end (\AA) & Transit depth (\%)                     &     $u_{\mathrm{1}}$ &      $u_{\mathrm{2}}$ &     $u_{\mathrm{3}}$ &      $u_{\mathrm{4}}$ \\
\hline
        2898 &      3700 & 1.050 $^{+0.037}$ $_{-0.044}$ & 0.5188 & -0.8368 & 2.0778 & -0.815  \\
        3700 &      4041 & 1.048 $^{+0.018}$ $_{-0.021}$ & 0.7102 & -1.1242 & 2.1235 & -0.7717 \\
        4041 &      4261 & 1.027 $^{+0.014}$ $_{-0.015}$ & 0.5224 & -0.6401 & 1.7752 & -0.738  \\
        4261 &      4426 & 1.028 $^{+0.014}$ $_{-0.013}$ & 0.6179 & -0.7511 & 1.7207 & -0.6878 \\
        4426 &      4536 & 1.025 $^{+0.009}$ $_{-0.012}$ & 0.4502 & -0.2005 & 1.2133 & -0.5686 \\
        4536 &      4646 & 1.035 $^{+0.016}$ $_{-0.018}$ & 0.4359 & -0.1007 & 1.0827 & -0.531  \\
        4646 &      4756 & 1.013 $^{+0.015}$ $_{-0.013}$ & 0.4582 & -0.146  & 1.1212 & -0.5536 \\
        4756 &      4921 & 1.023 $^{+0.011}$ $_{-0.011}$ & 0.4853 & -0.143  & 1.0708 & -0.5453 \\
        4921 &      5030 & 1.028 $^{+0.012}$ $_{-0.012}$ & 0.519  & -0.2152 & 1.0986 & -0.5414 \\
        5030 &      5140 & 1.034 $^{+0.016}$ $_{-0.015}$ & 0.5397 & -0.2945 & 1.1957 & -0.5839 \\
        5140 &      5250 & 1.005 $^{+0.014}$ $_{-0.012}$ & 0.6078 & -0.465  & 1.3201 & -0.616  \\
        5250 &      5360 & 1.024 $^{+0.019}$ $_{-0.014}$ & 0.5829 & -0.3203 & 1.1244 & -0.5466 \\
        5360 &      5469 & 1.024 $^{+0.019}$ $_{-0.016}$ & 0.575  & -0.2891 & 1.0862 & -0.5385 \\
        5469 &      5579 & 1.007 $^{+0.016}$ $_{-0.017}$ & 0.5927 & -0.3243 & 1.1044 & -0.5449 \\
        5579 &      5688 & 1.036 $^{+0.014}$ $_{-0.015}$ & 0.6041 & -0.3355 & 1.09   & -0.5378 \\
\hline
\end{tabular}
    \caption{Results from spectroscopic light curve fits for WASP-127b, using HST/STIS+G430L.  Fixed, four-parameter limb darkening law coefficients denoted by $u_{\mathrm{i}}$}
    \label{tab:w127_g430l}
\end{table*}

\begin{table*}
\centering
\begin{tabular}{cccccccc}
\hline
   Bin start (\AA) &   Bin end (\AA) & Transit depth (\%)                     &     $u_{\mathrm{1}}$ &      $u_{\mathrm{2}}$ &     $u_{\mathrm{3}}$ &      $u_{\mathrm{4}}$ \\
\hline
        5500 &      5600 & 1.023 $^{+0.027}$ $_{-0.021}$ & 0.5963 & -0.3281 & 1.1005 & -0.5429 \\
        5600 &      5700 & 1.047 $^{+0.022}$ $_{-0.023}$ & 0.6029 & -0.3282 & 1.0782 & -0.5331 \\
        5700 &      5800 & 1.014 $^{+0.014}$ $_{-0.014}$ & 0.5987 & -0.2994 & 1.0337 & -0.5182 \\
        5800 &      5878 & 1.051 $^{+0.018}$ $_{-0.020}$ & 0.593  & -0.2704 & 0.9903 & -0.5042 \\
        5878 &      5913 & 1.066 $^{+0.024}$ $_{-0.023}$ & 0.6326 & -0.3954 & 1.1143 & -0.5468 \\
        5913 &      6070 & 1.026 $^{+0.014}$ $_{-0.017}$ & 0.6126 & -0.3129 & 1.0114 & -0.5113 \\
        6070 &      6200 & 1.028 $^{+0.020}$ $_{-0.016}$ & 0.6537 & -0.4235 & 1.0947 & -0.5342 \\
        6200 &      6300 & 1.022 $^{+0.018}$ $_{-0.020}$ & 0.6632 & -0.4382 & 1.094  & -0.5331 \\
        6300 &      6450 & 1.036 $^{+0.015}$ $_{-0.016}$ & 0.6696 & -0.455  & 1.0978 & -0.5349 \\
        6450 &      6600 & 0.995 $^{+0.020}$ $_{-0.017}$ & 0.6986 & -0.4715 & 1.0601 & -0.5246 \\
        6600 &      6800 & 1.004 $^{+0.013}$ $_{-0.012}$ & 0.6883 & -0.5005 & 1.1059 & -0.5346 \\
        6800 &      7000 & 0.997 $^{+0.016}$ $_{-0.012}$ & 0.7015 & -0.5355 & 1.1128 & -0.5306 \\
        7000 &      7200 & 1.009 $^{+0.013}$ $_{-0.014}$ & 0.732  & -0.6214 & 1.1791 & -0.551  \\
        7200 &      7450 & 1.018 $^{+0.011}$ $_{-0.011}$ & 0.7449 & -0.658  & 1.1933 & -0.5537 \\
        7450 &      7645 & 1.003 $^{+0.017}$ $_{-0.016}$ & 0.7339 & -0.6262 & 1.1383 & -0.5317 \\
        7645 &      7720 & 1.020 $^{+0.022}$ $_{-0.026}$ & 0.7356 & -0.6251 & 1.115  & -0.5196 \\
        7720 &      8100 & 1.010 $^{+0.020}$ $_{-0.019}$ & 0.749  & -0.6699 & 1.1521 & -0.5322 \\
        8100 &      8485 & 0.986 $^{+0.026}$ $_{-0.018}$ & 0.7713 & -0.7465 & 1.2143 & -0.555  \\
        8485 &      8985 & 1.005 $^{+0.023}$ $_{-0.018}$ & 0.7557 & -0.7033 & 1.1225 & -0.5143 \\
        8985 &     10240 & 1.025 $^{+0.018}$ $_{-0.017}$ & 0.7199 & -0.6086 & 0.9678 & -0.4452 \\
\hline
\end{tabular}
    \caption{Results from spectroscopic light curve fits for WASP-127b, using HST/STIS+G750L.  Fixed, four-parameter limb darkening law coefficients denoted by $u_{\mathrm{i}}$}
    \label{tab:w127_g750l}
\end{table*}

\begin{table*}
\centering
\begin{tabular}{cccccccc}
\hline
   Bin start (\AA) &   Bin end (\AA) & Transit depth (\%)                     &     $u_{\mathrm{1}}$ &      $u_{\mathrm{2}}$ &     $u_{\mathrm{3}}$ &      $u_{\mathrm{4}}$ \\
\hline
       11225 &     11409 & 0.993 $^{+0.009}$ $_{-0.009}$ & 0.6515 & -0.4064 &  0.6627 & -0.3206 \\
       11409 &     11594 & 0.995 $^{+0.008}$ $_{-0.007}$ & 0.6312 & -0.3375 &  0.5764 & -0.2851 \\
       11594 &     11779 & 1.001 $^{+0.009}$ $_{-0.010}$ & 0.6281 & -0.3166 &  0.544  & -0.2745 \\
       11779 &     11963 & 0.991 $^{+0.010}$ $_{-0.009}$ & 0.6205 & -0.287  &  0.5009 & -0.2564 \\
       11963 &     12148 & 1.001 $^{+0.008}$ $_{-0.009}$ & 0.6093 & -0.2483 &  0.4502 & -0.2362 \\
       12148 &     12333 & 0.991 $^{+0.008}$ $_{-0.008}$ & 0.5884 & -0.162  &  0.3459 & -0.1968 \\
       12333 &     12517 & 0.992 $^{+0.008}$ $_{-0.007}$ & 0.5787 & -0.1206 &  0.2867 & -0.1716 \\
       12517 &     12702 & 1.000 $^{+0.012}$ $_{-0.011}$ & 0.5727 & -0.0874 &  0.2388 & -0.1533 \\
       12702 &     12887 & 0.993 $^{+0.008}$ $_{-0.008}$ & 0.5709 & -0.0386 &  0.1613 & -0.1299 \\
       12887 &     13071 & 0.997 $^{+0.009}$ $_{-0.009}$ & 0.554  &  0.006  &  0.1075 & -0.0997 \\
       13071 &     13256 & 1.010 $^{+0.007}$ $_{-0.008}$ & 0.5476 &  0.0484 &  0.0384 & -0.0685 \\
       13256 &     13441 & 1.016 $^{+0.008}$ $_{-0.008}$ & 0.5386 &  0.1046 & -0.0462 & -0.0321 \\
       13441 &     13625 & 1.054 $^{+0.009}$ $_{-0.010}$ & 0.5338 &  0.1452 & -0.1094 & -0.0057 \\
       13625 &     13810 & 1.051 $^{+0.011}$ $_{-0.011}$ & 0.5332 &  0.1813 & -0.1788 &  0.0266 \\
       13810 &     13995 & 1.052 $^{+0.008}$ $_{-0.008}$ & 0.5265 &  0.2444 & -0.2789 &  0.0708 \\
       13995 &     14179 & 1.046 $^{+0.007}$ $_{-0.007}$ & 0.5238 &  0.2836 & -0.3521 &  0.1059 \\
       14179 &     14364 & 1.053 $^{+0.007}$ $_{-0.006}$ & 0.5301 &  0.2999 & -0.4012 &  0.1308 \\
       14364 &     14549 & 1.046 $^{+0.010}$ $_{-0.011}$ & 0.5418 &  0.3015 & -0.431  &  0.1464 \\
       14549 &     14733 & 1.042 $^{+0.009}$ $_{-0.009}$ & 0.5518 &  0.3122 & -0.4668 &  0.1632 \\
       14733 &     14918 & 1.031 $^{+0.009}$ $_{-0.010}$ & 0.567  &  0.29   & -0.4699 &  0.1709 \\
       14918 &     15102 & 1.028 $^{+0.009}$ $_{-0.012}$ & 0.5795 &  0.2891 & -0.5072 &  0.1952 \\
       15102 &     15287 & 1.012 $^{+0.010}$ $_{-0.010}$ & 0.5983 &  0.31   & -0.585  &  0.2369 \\
       15287 &     15472 & 0.999 $^{+0.010}$ $_{-0.009}$ & 0.631  &  0.2627 & -0.5741 &  0.2409 \\
       15472 &     15656 & 1.001 $^{+0.010}$ $_{-0.011}$ & 0.6489 &  0.2307 & -0.5607 &  0.2408 \\
       15656 &     15841 & 0.983 $^{+0.009}$ $_{-0.009}$ & 0.6836 &  0.13   & -0.4668 &  0.2097 \\
       15841 &     16026 & 0.986 $^{+0.009}$ $_{-0.009}$ & 0.7076 &  0.0634 & -0.4054 &  0.19   \\
       16026 &     16210 & 0.975 $^{+0.012}$ $_{-0.011}$ & 0.7347 &  0.0371 & -0.4274 &  0.21   \\
       16210 &     16395 & 0.967 $^{+0.013}$ $_{-0.012}$ & 0.7468 &  0.0085 & -0.4018 &  0.2017 \\
\hline
\end{tabular}
    \caption{Results from spectroscopic light curve fits for WASP-127b, using HST/WFC3+G141.  Fixed, four-parameter limb darkening law coefficients denoted by $u_{\mathrm{i}}$}
    \label{tab:w127_wfc3}
\end{table*}

\begin{figure}
\begin{minipage}[b]{.47\textwidth}
\centering
\includegraphics[width=\linewidth]{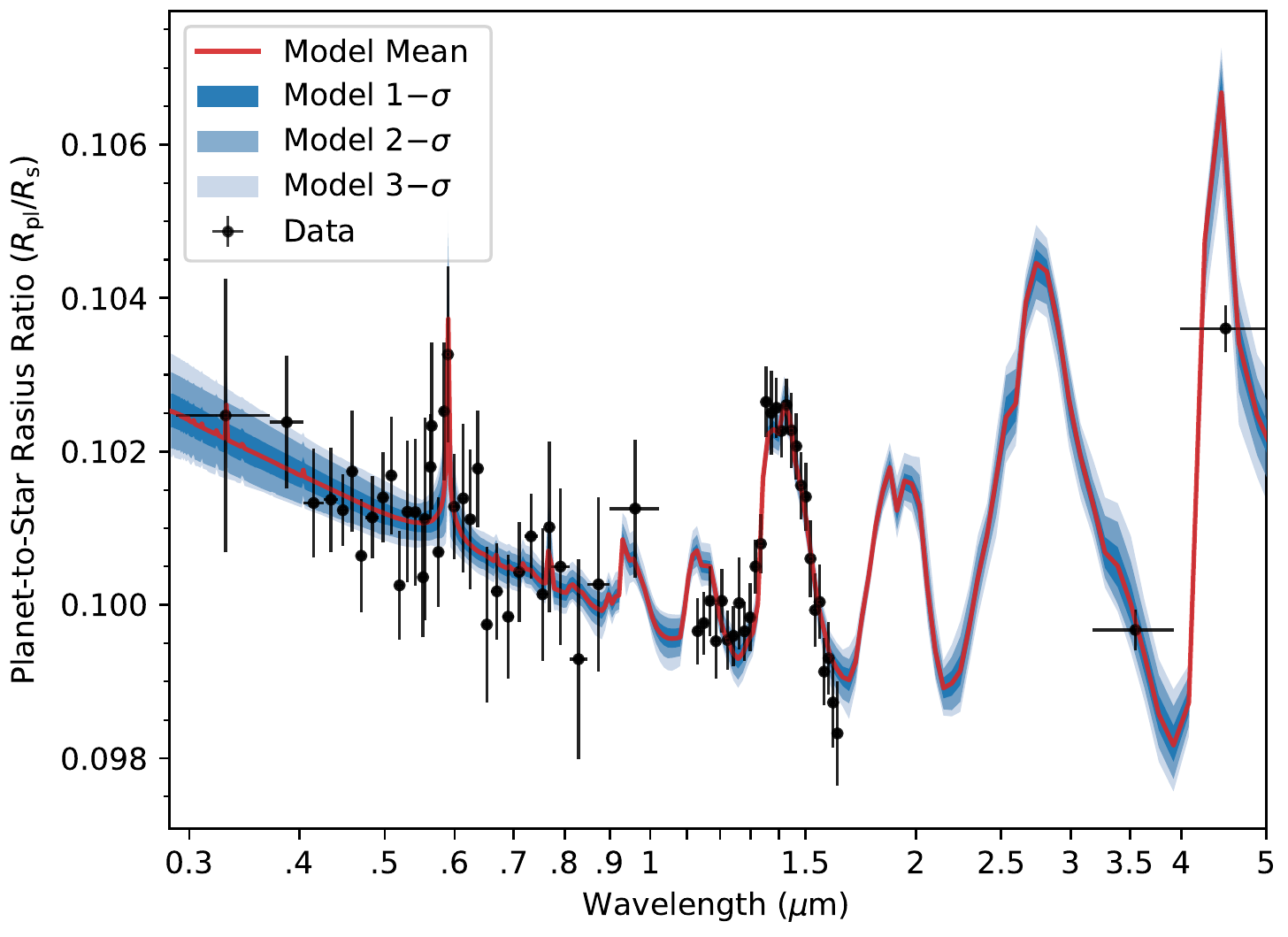}
\end{minipage}
\hfill
\begin{minipage}[b]{.47\textwidth}
\centering
\includegraphics[width=\linewidth]{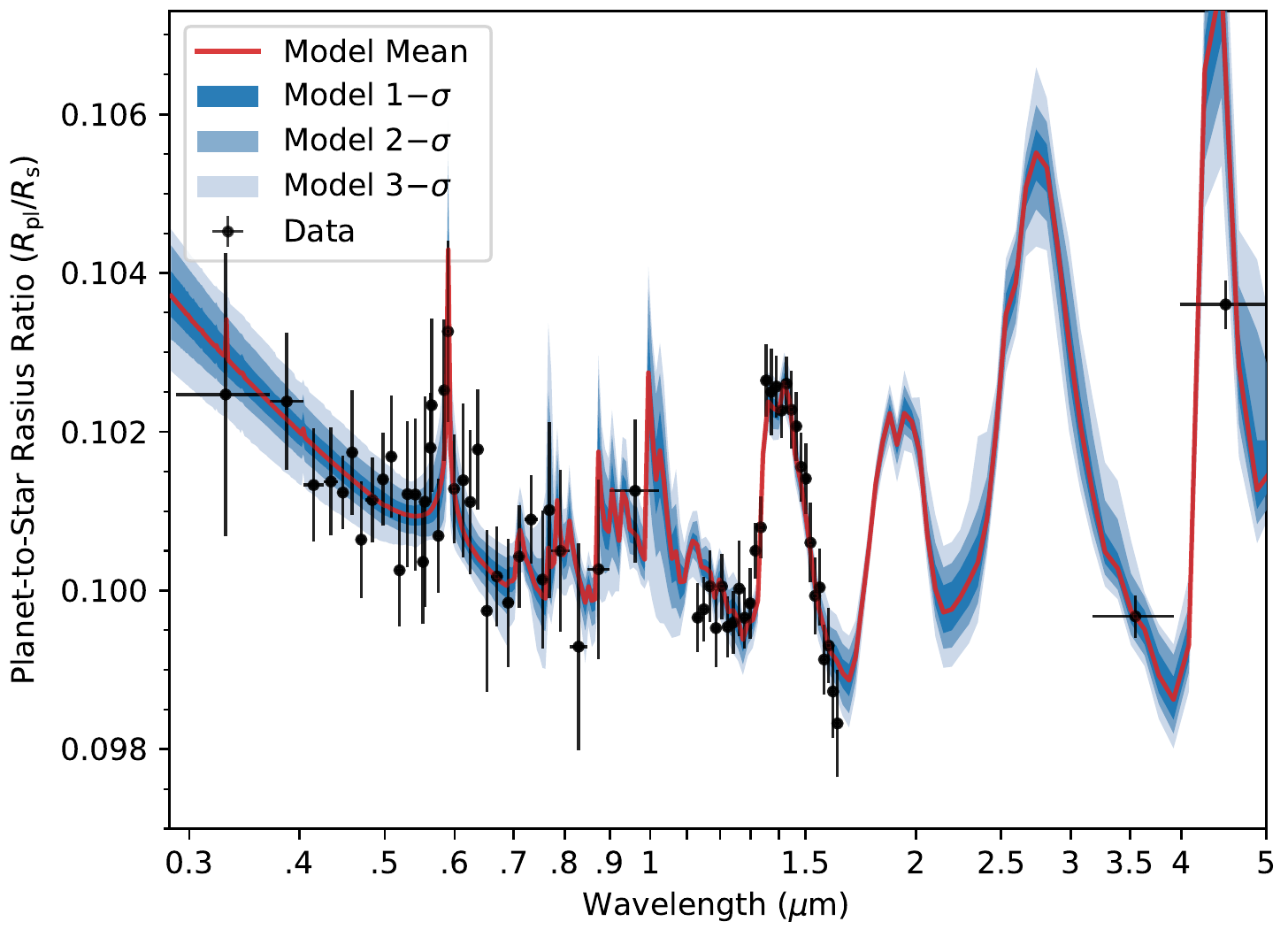}
\vspace*{-1\baselineskip}          %Use command to remove white space
\caption{Transmission spectrum retrieval results from ATMO assuming chemical equilibrium (top) and free chemistry (bottom). The black points are the data, the red line is the median spectrum of the posterior distribution. The 1, 2, and 3-$\sigma$ model distributions are also shown (dark blue, blue, light blue respectively).}\label{fig:w127_spectra_retrievals}
\end{minipage}
%\vspace*{-1\baselineskip}          %Use command to remove white space
\end{figure}

\begin{figure}
\begin{minipage}[b]{.47\textwidth}
\centering
\includegraphics[width=\linewidth]{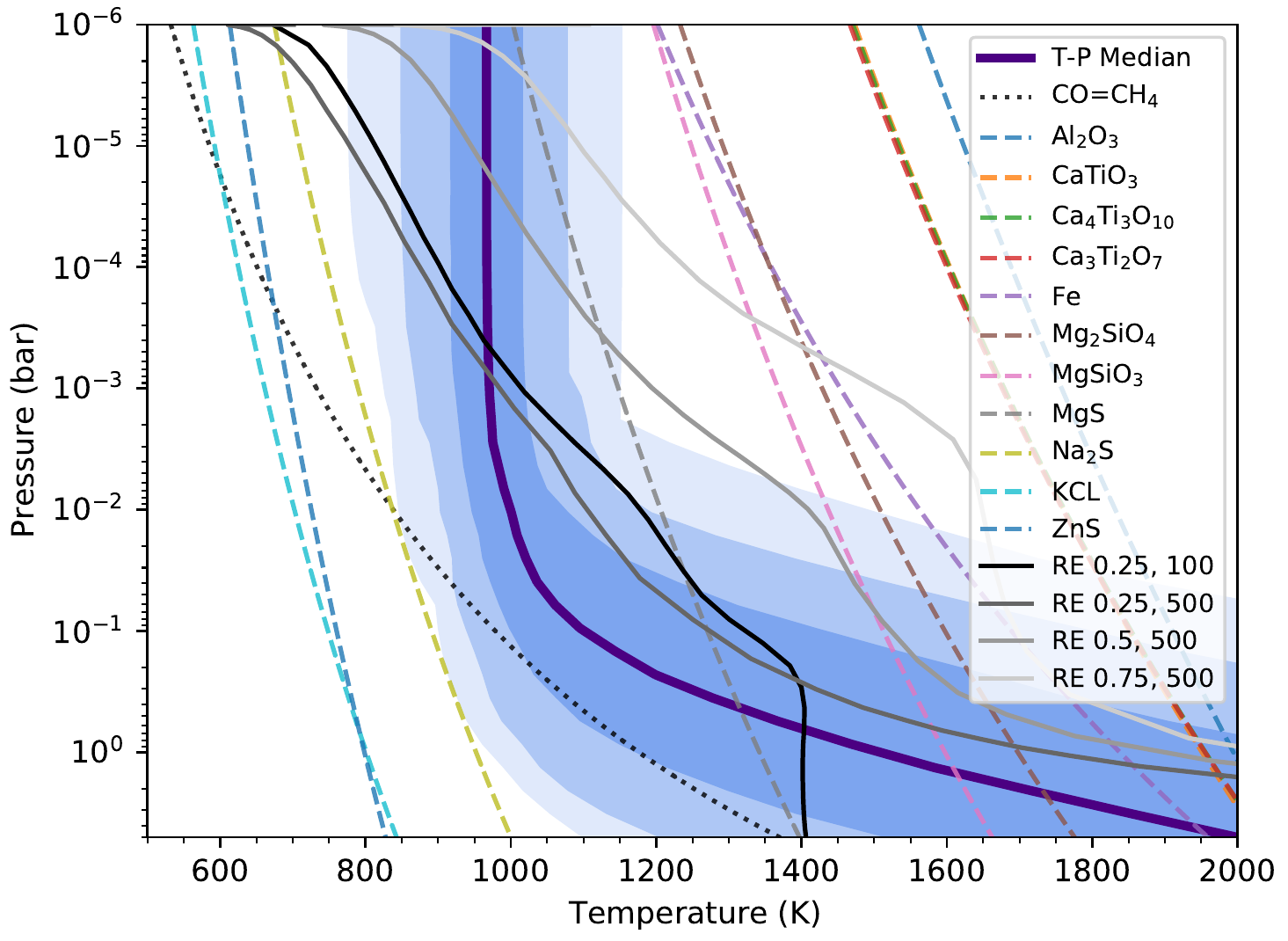}
\end{minipage}
\hfill
\begin{minipage}[b]{.47\textwidth}
\centering
\includegraphics[width=\linewidth]{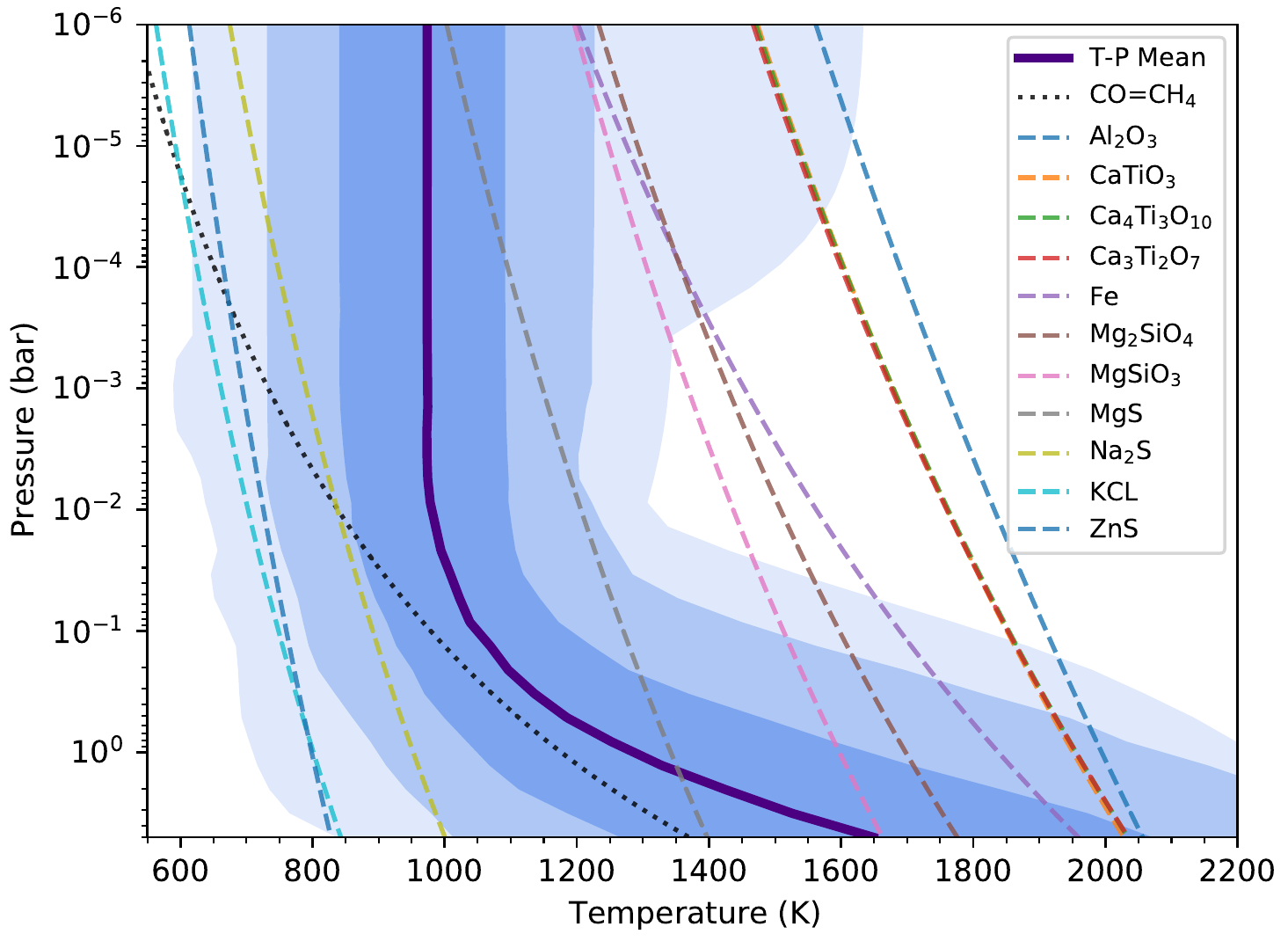}
%\vspace*{-1\baselineskip}          %Use command to remove white space
\caption{Pressure-Temperature profile retrieval results from ATMO assuming chemical equilibrium (top) and free chemistry (bottom). The solid purple line is the median profile of the posterior distribution; the dark blue, blue, and light blue shaded regions are the 1, 2, and 3-$\sigma$ model distributions. The dashed lines are cloud condensation curves calculated at a metallicity of 10$\times$ solar from \protect\cite{2006ApJ...648.1181V,2010ApJ...716.1060V} and \protect\cite{2017MNRAS.464.4247W}.  The three grey solid lines in the top panel are converged P-T profiles in chemical- and radiative equilibrium (RE), each with an internal temperature ($T_{int}$) of 500K, and a heat recirculation factor ($f_{c}$) of 0.25, 0.5, and 0.75 respectively.  They use the haze and elemental abundances from the chemical equilibrium retrieval. Black solid line is a cloud-free converged T-P profile at 10$\times$ solar abundance for WASP-127b from \protect\cite{Goyal2020arXiv200801856G}, with $T_{int}$=100K and $f_{c}$=0.25. }
\label{fig:w127_PT_retrievals}
\end{minipage}
%\vspace*{-1\baselineskip}          %Use command to remove white space
\end{figure}

\section{ATMO Retrieval}
\label{sec:arc}
For the combined STIS, WFC3 and Spitzer spectrum of WASP-127b, we performed an atmospheric retrieval analysis using a one-dimensional radiative transfer code for planetary atmospheres, ATMO (\citealt{2014A&A...564A..59A, 2015ApJ...804L..17T, 2016ApJ...817L..19T, 2017ApJ...841...30T, 2018MNRAS.474.5158G}).  
Previous retrieval results using ATMO can be found in \cite{2017Sci...356..628W}, \cite{2017Natur.548...58E}, and \cite{2019MNRAS.488.2222M}.
ATMO can solve for the pressure-temperature (P-T) profile and chemical abundances that satisfy hydrostatic equilibrium and conservation of energy given a set of opacities (references for the opacity sources are listed in \citealt{2018MNRAS.474.5158G}).  It can also fit a parameterised P-T profile and chemical abundances to transmission spectroscopy data.  For our retrievals, we used the paramaterised P-T profile of \cite{2010A&A...520A..27G} which gives three free parameters: the Planck mean thermal infrared opacity $(\upkappa_{IR})$; the ratio of optical to infrared opacities $(\gamma_{O/IR})$; and an irradiation efficiency factor $(\beta)$.  The internal temperature was set to 500 K based on \cite{Thorngren2019ApJ...884L...6T}.
We used a relatively simple haze model: 
\begin{equation}
    \sigma (\lambda)_{\mathrm{haze}} = \delta_{\mathrm{haze}} \sigma_{\mathrm{0}}(\lambda/\lambda_0)^{-\alpha_{\mathrm{haze}}},
\end{equation}
where $\sigma (\lambda)$ is the total scattering cross-section of the haze; $\delta_{\mathrm{haze}}$ is an empirical enhancement factor; $\sigma_{\mathrm{0}}$ is the scattering cross section of molecular hydrogen at 0.35 $\mu$m; and $\alpha_{\mathrm{haze}}$ sets the wavelength dependence of the scattering.  For example, for pure Rayleigh scattering, $\alpha_{\mathrm{haze}}=4 $. 

We performed two retrievals, one in chemical equilibrium with the elemental abundances free to vary and a free chemistry model where the molecular abundances were freely fit.   

We coupled the forward \texttt{ATMO} model to a nested sampling algorithm
\citep{feroz2008,feroz2009,feroz2013}
enabling Bayesian model comparison and marginalizing the posterior distribution.

\begin{figure*}
    \centering
    \includegraphics[width=\textwidth]{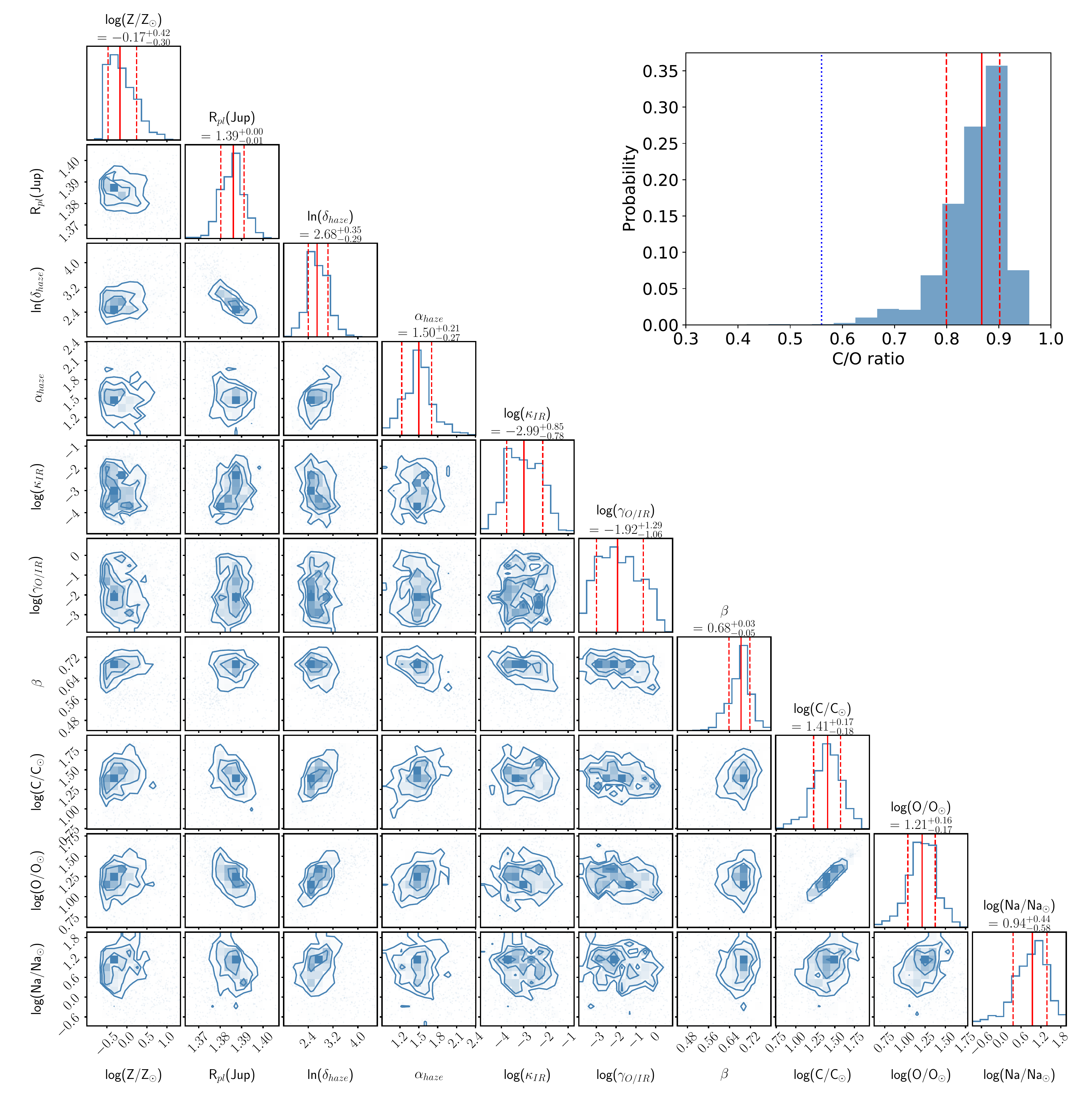}
    \vspace*{-1\baselineskip}          %Use command to remove white space
    \caption{Posterior distributions for atmospheric retrieval fit from ATMO for WASP-127b assuming chemical equilibrium. Solid red lines show median values; dashed red lines contain 68\% of samples.  The posterior distribution of the C/O ratio is also shown, with the solar value shown as the blue dotted line.}
    \label{fig:w127_eq_mcmc}
\end{figure*}

For the model assuming chemical equilibrium, the elemental abundances for each model were freely fit and calculated in equilibrium on the fly.  Five elements were selected to vary independently, as they are major species which are also likely to be sensitive to spectral features in the data, while the rest were varied by a trace metallicity parameter ([Z$_{\rm trace}$/Z$_{\odot}$]).  By varying the carbon, oxygen, and sodium  %and potassium
elemental abundances ([C/C\textsubscript{$\odot$}], [O/O\textsubscript{$\odot$}], [Na/Na\textsubscript{$\odot$}]) %, [K/K\textsubscript{$\odot$}], [Li/Li\textsubscript{$\odot$}])
separately we allow for non-solar compositions but with chemical equilibrium imposed such that each model fit has a chemically-plausible mix of molecules given the retrieved temperatures, pressures and underlying elemental abundances. Importantly, by varying both O and C separately (rather than a single C/O value) we alleviate an important modeling assumption which can affect the retrieved C/O value (see \citealt{2019MNRAS.486.1123D}). For the spectral synthesis, we included the spectroscopically active molecules of H$_{2}$, He, H$_{2}$O, CO$_{2}$, CO, CH$_{4}$, NH$_{3}$, H$_2$S, HCN, C$_2$H$_2$, Na, K, Li, TiO, VO, Fe, FeH, SO$_2$, HCN, H$_2$S, PH$_3$, and H-.  Rainout of condensate species was also included.  The results are shown in Figs. \ref{fig:w127_spectra_retrievals} and \ref{fig:w127_eq_mcmc} with the results also given in Table \ref{tab:arceq} including the best-fitting model parameters and 1 $\sigma$ uncertainties (which correspond to the range of parameters which contain 68\% of the posterior samples).  Good fits to the data were obtained, with the minimum $\chi^2$ of 47.6 found for 55 degrees of freedom.  We also calculated physically consistent P-T profiles in chemical- and radiative equilibrium using the best-fit elemental abundances and haze parameters from the chemical equilibrium retrieval (Figure \ref{fig:w127_PT_retrievals}).  For these numerically-simulated P-T profiles we varied the internal temperature of the planet, $T_{int}$, and its atmospheric heat recirculation factor, $f_{c}$. We reduce the incoming flux in the 1D column of the atmosphere by $f_{c}$ to account for the effects of winds which redistribute stellar flux around the 3-D planet, and the incidence angle of the flux. 

\begin{figure*}
    \centering
    \includegraphics[width=\textwidth]{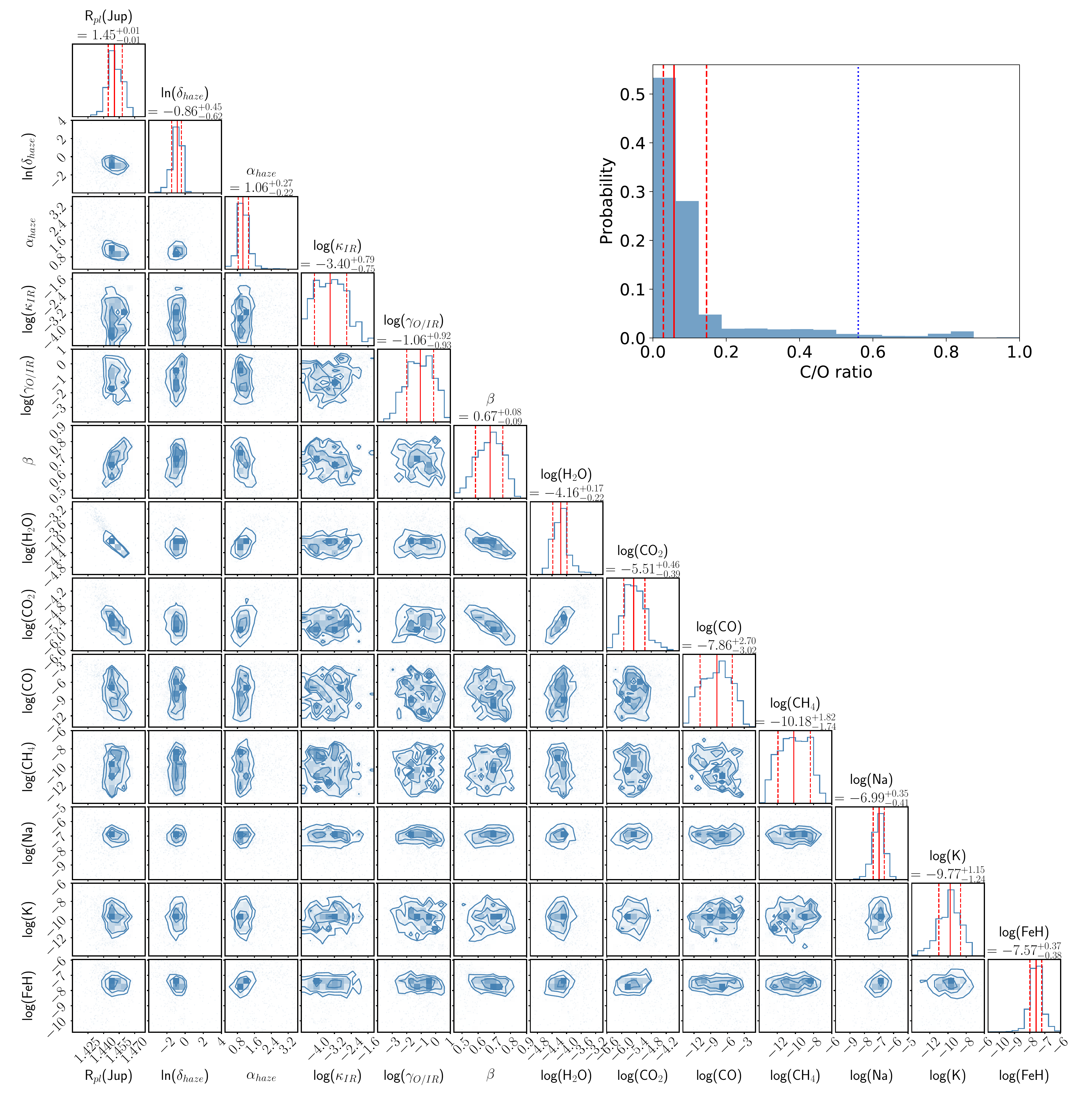}
    \vspace*{-1\baselineskip}          %Use command to remove white space
    \caption{Similar as Fig. \ref{fig:w127_eq_mcmc} but for freely fit chemistry.}
    \label{fig:w127_free_mcmc}
\vspace*{-1\baselineskip}          %Use command to remove white space
\end{figure*}

In the free-chemistry model, we assumed that the volume mixing ratio (VMR) for each species was constant with pressure, and each molecule VMR was independently fit.  We varied the H\textsubscript{2}O, CO\textsubscript{2}, CO, CH$_{4}$, Na, K, and FeH abundances.  Li was not fit nor was TiO, VO, HCN, Fe, and C$_2$H$_2$ as no signs of them were observed in the data.  The scattering haze scheme was the same as the equilibrium model.  The results are shown in Figs. \ref{fig:w127_spectra_retrievals}, \ref{fig:w127_free_mcmc} with the results also given in Table \ref{tab:arcfree}.   The free-retrieval also resulted in a good fit, with a minimum $\chi^2$ of 42.7 found for 52 degrees of freedom.

We ran two additional retrievals to test the influence of our choice of scattering haze scheme on our final results.  First, we re-ran the chemical equilibrium model without any haze (i.e. a 'clear' atmosphere).  The best-fit model gave an extremely poor fit to the data, with a minimum $\chi^2$ of 264 for 57 degrees of freedom.  In particular, it could not produce enough opacity shortwards of 0.6 $\mu$m to match the data; and in order to match the muted water feature at 1.4 $\mu$m, it fit a much cooler P-T profile (500 K at pressures of 10$^{-2} - 10^{-6}$ bar, which seems implausible for a planet with an equilibrium temperature of 1400 K).  We note that clear atmosphere solutions were allowed in all of our retrievals, since the uniform priors on the haze parameters encompass no-haze scenarios, but they were disfavoured.  We therefore conclude that WASP-127b's atmosphere is unlikely to be entirely haze- or cloud-free.  

Secondly, we ran the 'hazy' equilibrium chemistry retrieval with an additional, condensate `cloud' absorption, which was assumed to have a grey wavelength dependence, and was calculated as 
\begin{equation}
    \kappa(\lambda)_{\mathrm{cloud}} = \delta_{\mathrm{cloud}}\kappa_{\mathrm{H2}}, 
\end{equation}
where $\kappa(\lambda)_{\mathrm{c}}$ is the 'cloud' absorption opacity, $\delta_{\mathrm{cloud}}$ is an empirical factor governing the strength of the grey scattering, and $\kappa_{\mathrm{H2}}$ is the scattering opacity due to H$_{2}$ at 0.35 $\mu$m. The previously-described haze, $\sigma (\lambda)_{\mathrm{haze}}$, and the cloud, $\kappa(\lambda)_{\mathrm{cloud}}$, were added to the total gaseous scattering and absorption respectively.  Our best-fit result from the haze-plus-cloud run provides an equally good fit to the data as the haze-only scheme, with a minimum $\chi^2$ of 47.7 found for 53 degrees of freedom.  The best-fit spectrum is almost identical to the haze-only run (see Figure \ref{fig:w127_spectra_retrievals}), because the data favours a very low level of wavelength-independent absorption.  The retrieved molecular abundances all fall within 1-$\sigma$ of the haze-only retrieved abundances.  We therefore conclude that there is little evidence in the data for a grey absorbing cloud.

\begin{figure}
    \centering
    \includegraphics[width=\columnwidth]{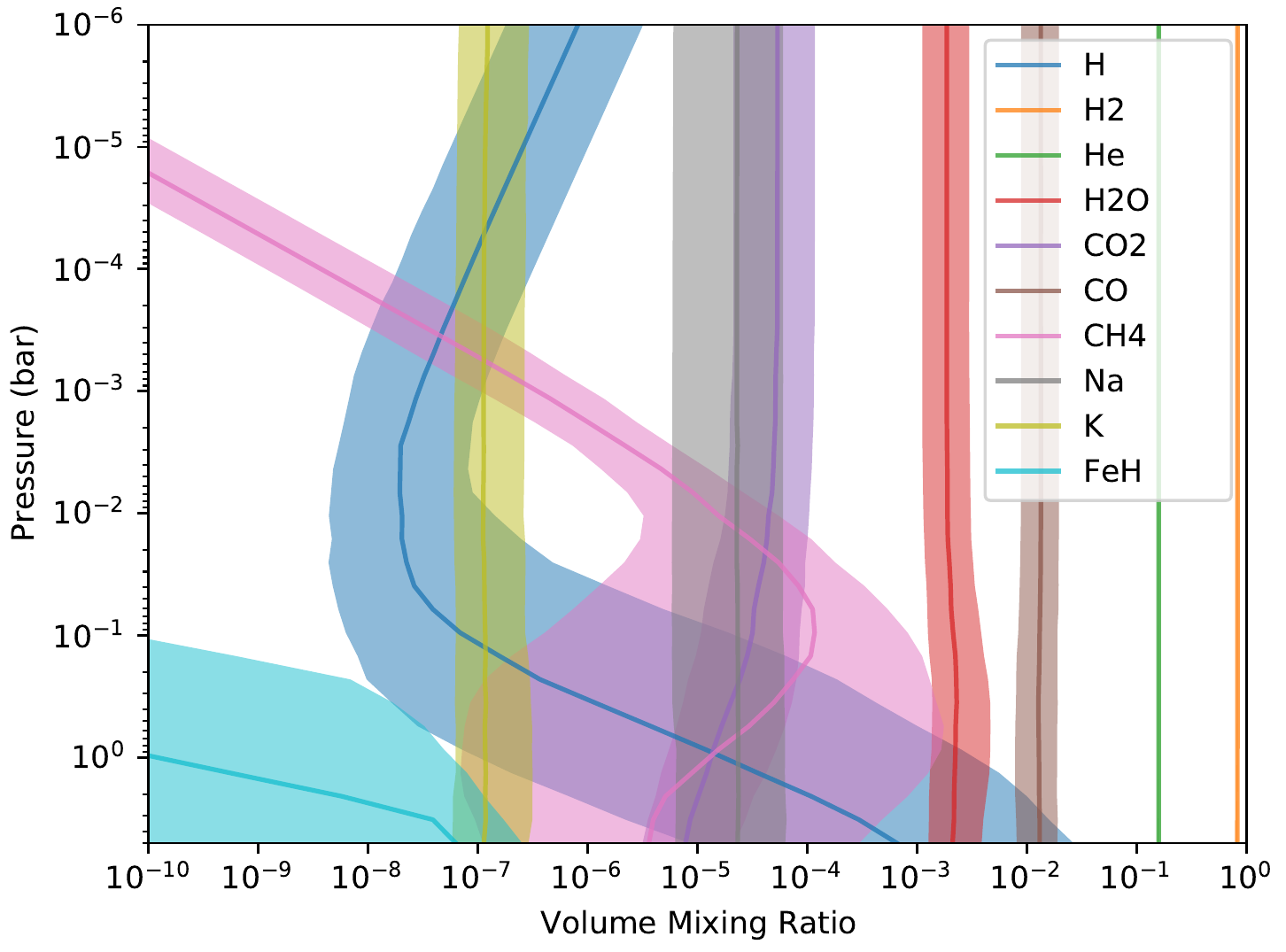}
    \vspace*{-1\baselineskip}          %Use command to remove white space
    \caption{Retrieved chemical abundances as a function of pressure from the ATMO chemical equilibrium retrieval, for selected species.}
    \label{fig:AMOchem}
\vspace*{-1\baselineskip}          %Use command to remove white space
\end{figure}

\section{NEMESIS Retrieval}
We also performed an independent retrieval using the NEMESIS radiative transfer and retrieval tool. NEMESIS was originally developed for Solar System analysis \citep{irwin08} and subsequently updated for the study of exoplanets  (e.g. \citealt{Lee12, barstow17}). More recent upgrades have included incorporating the PyMultiNest nested sampling algorithm \citep{krissansen-totton_detectability_2018, feroz2008, feroz2009, feroz2013, Buchner2014}. NEMESIS uses the correlated-k approximation for tabulating atomic and molecular opacities \citep{lacis91}, which results in a rapid forward model computation. 
NEMESIS incorporates absorption from Na, K, H$_{2}$O, CO$_{2}$, CO and CH$_{4}$, as well as collision-induced absorption due to H$_{2}$ and He.  Sources of absorption line data are included in Table \ref{tab:nemesis_opacities}. K tables used in this work are as compiled in Chubb et al. (Submitted). The retrieved temperature profile follows the specification of \cite{2010A&A...520A..27G}, and the haze scheme is the same as that described in \ref{sec:arc}, except we allow for incomplete ('patchy') coverage, as described in \cite{2017MNRAS.469.1979M}.

\begin{table}
    \centering
    \begin{tabular}{lp{.3\textwidth}}
    \hline
    Opacity source & Reference \\
    \hline
    CO & \cite{colinelist} \\
    CO$_{2}$ & \cite{co2linelist} \\
    CH$_{4}$ & \cite{ch4linelist} \\
    H$_{2}$O & \cite{h2o_linelist} \\
    H$_{2}$ - H$_{2}$ & \cite{borysow02}; \par \cite{borysow90} \\
    H$_{2}$ - He & \cite{borysow89}; \par \cite{borysowfm89} \\
    Na & \cite{NaKlinelist} \\
    K & \cite{NaKlinelist} \\
    \hline
    \end{tabular}
\caption{ References for opacity sources used in NEMESIS retrieval. }
\label{tab:nemesis_opacities}
\end{table}

\begin{figure}
    \centering
    \includegraphics[width=\columnwidth]{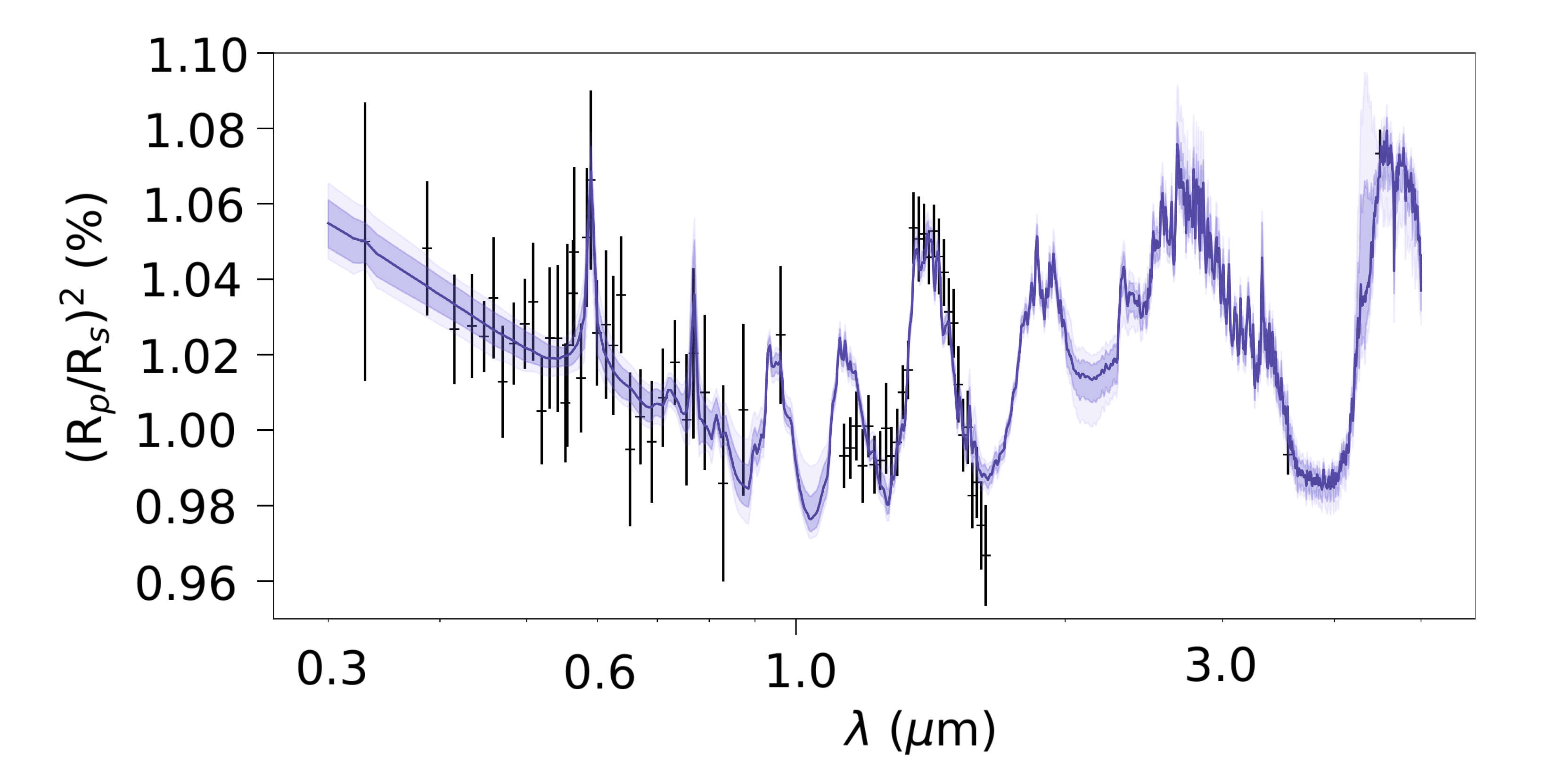}
    \vspace*{-1\baselineskip}          %Use command to remove white space
    \caption{Transmission spectrum retrieval results from NEMESIS}
    \label{fig:nemesis_spec}
\vspace*{-1\baselineskip}          %Use command to remove white space
\end{figure}

\begin{figure}
    \centering
    \includegraphics[width=\columnwidth]{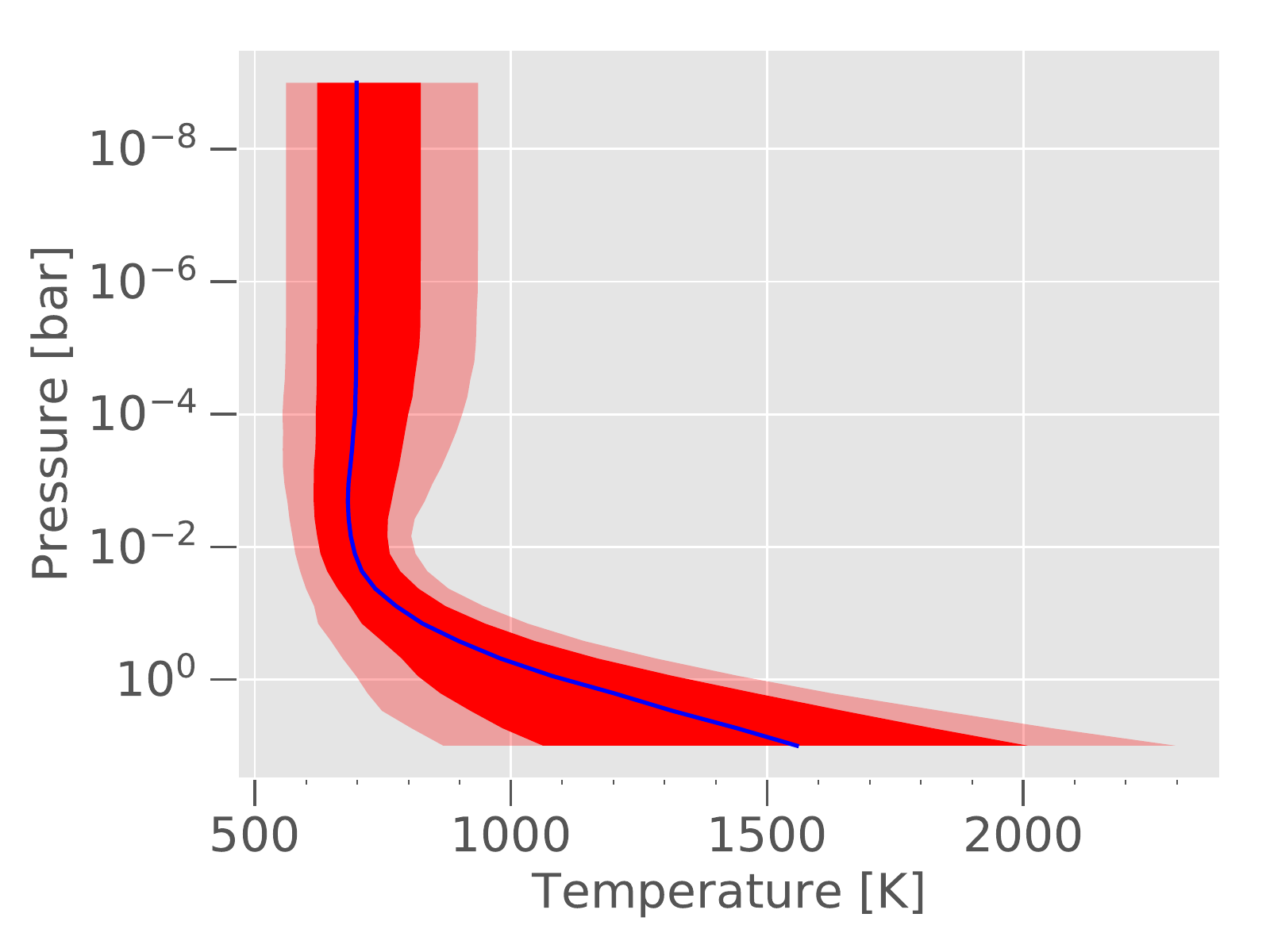}
    \vspace*{-1\baselineskip}          %Use command to remove white space
    \caption{Pressure-Temperature profile retrieval results from NEMESIS}
    \label{fig:nemesis_pt}
\vspace*{-1\baselineskip}          %Use command to remove white space
\end{figure}

\begin{figure*}
    \centering
    \includegraphics[width=\textwidth]{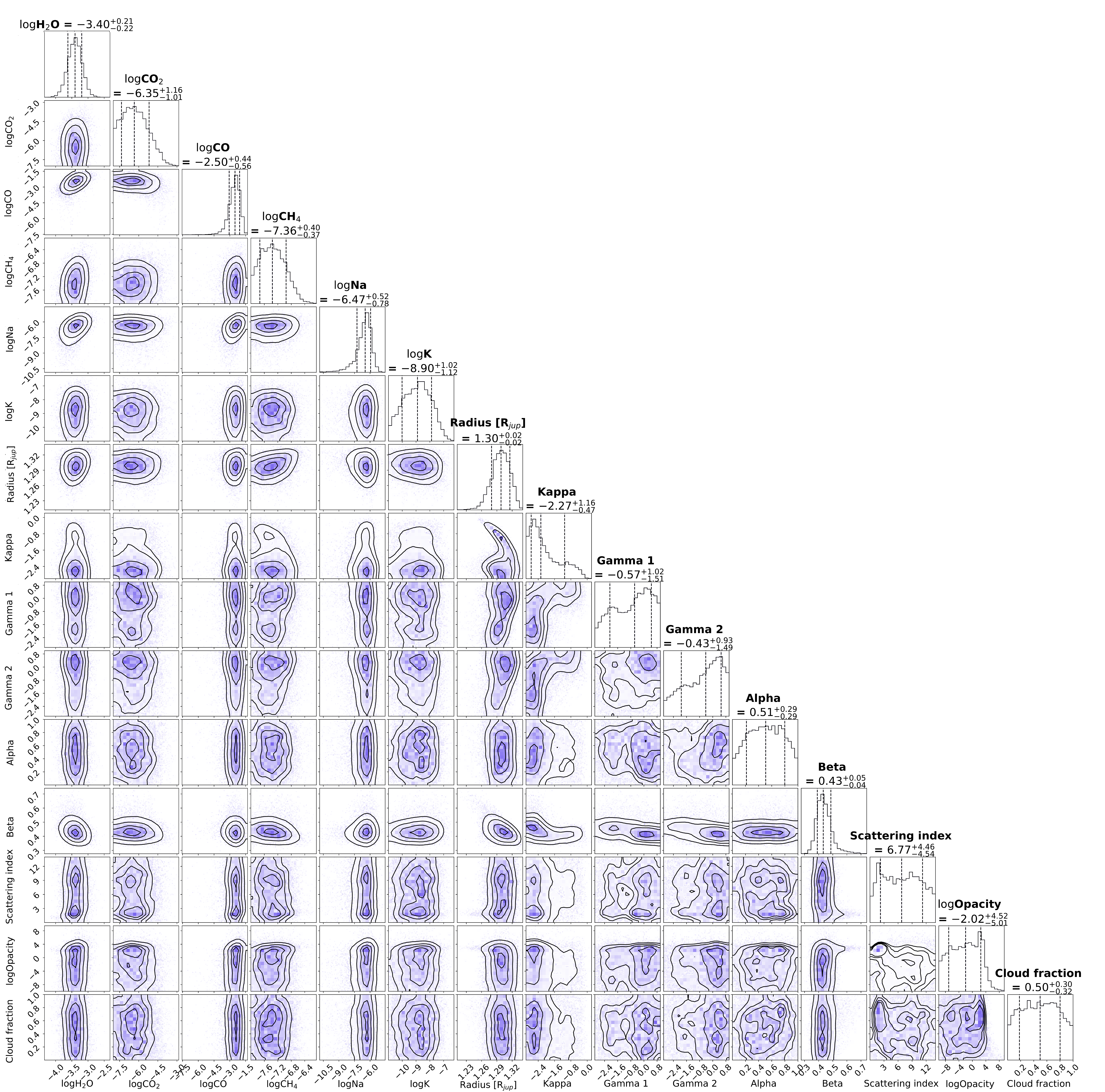}
    \vspace*{-1\baselineskip}          %Use command to remove white space
    \caption{Posterior distributions for atmospheric retrieval fit from NEMESIS for WASP-127b}
    \label{fig:nemesis_triangle}
\end{figure*}

\section{Results and Discussion}
\begin{figure*}
    \centering
    \includegraphics[width=\textwidth]{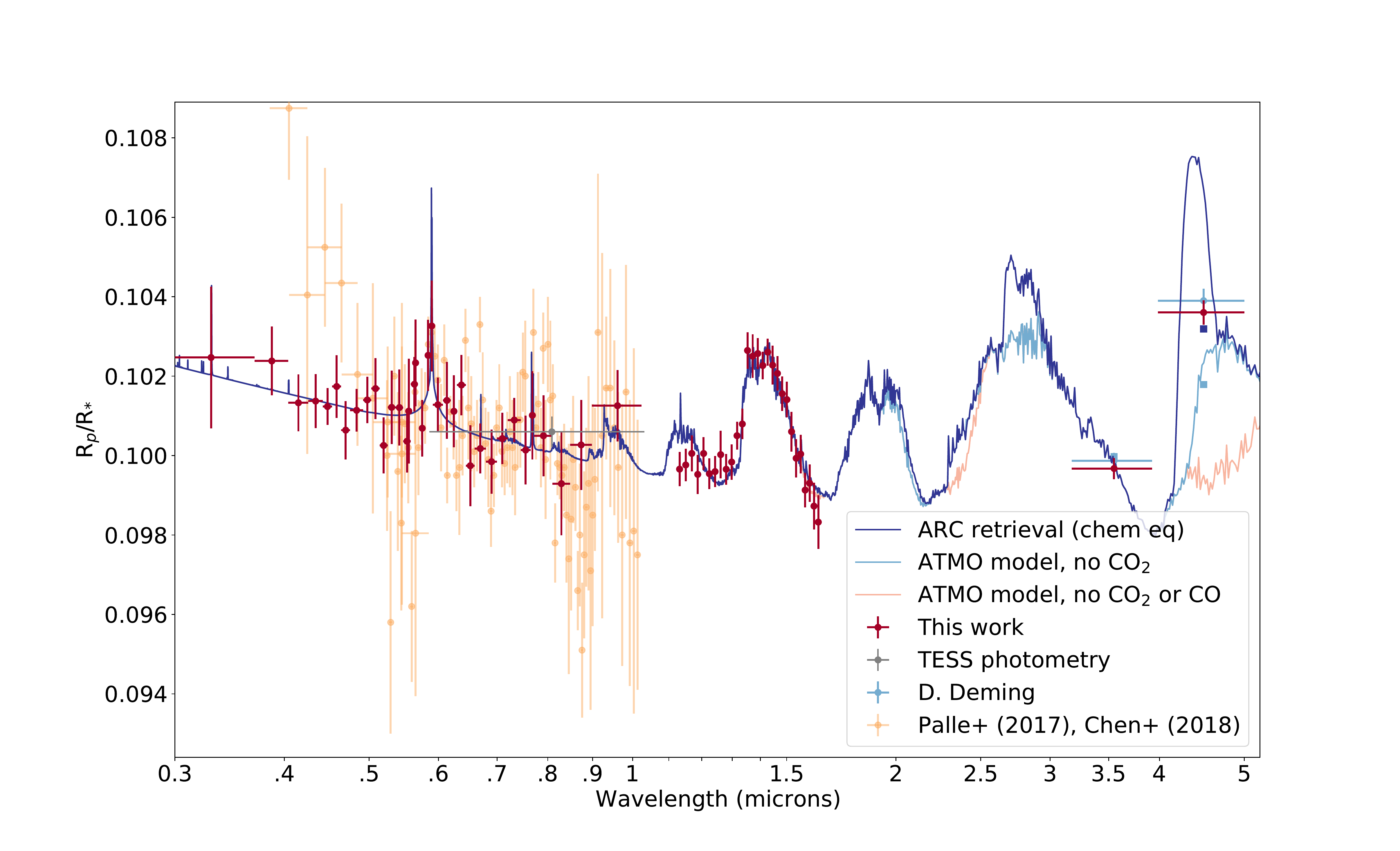}
    \vspace*{-1\baselineskip}          %Use command to remove white space
    \caption{Transmission spectrum for WASP-127b.  Red data points are from this work, blue data points are from an independent analysis of the Spitzer light curves from D. Deming.  Yellow points are previously published data from Palle et al. (2017) and Chen et al (2018) %\cite{2017A&A...602L..15P} and \cite{2018A&A...616A.145C} 
    using ground-based telescopes.  Dark blue line is are our best-fit retrieval models using ARC and MPFIT, light blue line is same model with CO$_{2}$ abundance set to zero. Square points show models binned to the resolution of the Spitzer data.}
    \label{fig:w127_spec}
\end{figure*}

The transmission spectrum of WASP-127b shows absorption by sodium, water, and either CO$_{2}$ or CO (see Figure \ref{fig:w127_spec}).  Longward of $\sim$0.5 $\mu$m, the optical spectrum is in good agreement with the results of Chen et al. (2018), albeit with a few points which differ with a statistically significant margin.  To quantify the detection significances of sodium and water, we calculated the Bayesian Information Criterion (BIC) for our best-fit ATMO free-chemistry model (BIC$_{i,model}$); and a straight line fit through the data range of interest (BIC$_{i,line}$).  For sodium, we used all HST/STIS data at wavelengths shorter than 0.722 $\mu$m, because at these wavelengths our models suggest the only opacity sources are sodium and a scattering haze.  For these 28 data points, we found BIC$_{Na,model}$ = 22.9 and BIC$_{Na,line}$ = 27.9, with 3 and 2 free parameters, respectively.  The ATMO model provides a better fit to the data at 1.7 $\sigma$, which is a detection of low significance, compared to the previous detection of sodium at 4.1$\sigma$ for WASP-127b by Chen et al. (2018).  For water, we used all of the data from HST/WFC3, and found BIC$_{H_{2}O,model}$ = 35.9 and BIC$_{H_{2}O,line}$ = 228.9, which gives a detection significance of 13.7 $\sigma$.  For context, one of the strongest water absorption features measured in an exoplanet atmosphere, also using a single HST/WFC3 transit observation in spatial scan mode, was made by \cite{2018ApJ...858L...6K} for WASP-107b - a similarly low-density gas-giant exoplanet, but cooler than WASP-127b.  They detected water in WASP-107b's atmosphere at a confidence of 6.5 $\sigma$.  To the best of our knowledge, the dataset presented here shows the strongest water absorption feature (in terms of detection significance) of any exoplanet observed with HST/WFC3. 

The sodium and water abundances are well-constrained by our ATMO retrievals, and are super-solar.  Our free-chemistry fit gives a water abundance of -4.16$^{+0.17}_{-0.22}$, which is consistent with, but better-constrained than the value of log$(H_{2}O) = -2.50^{+0.94}_{-4.56}$ found by \cite{2018A&A...616A.145C}, who observed unresolved water absorption.  We retrieved a lower sodium abundance of -6.99$^{+0.35}_{-0.41}$, compared to the very super-solar value of log($Na$) = $-3.17^{+1.03}_{-1.46}$) from \cite{2018A&A...616A.145C}  ($log_{10}[Na_{\odot}] = -5.76$, Asplund et al. 2009).  The NEMESIS retrieval gives water and sodium abundances of $-3.40^{+0.21}_{-0.22}$ and $-6.47^{+0.52}_{-0.78}$ respectively, which both agree with the ATMO abundances at the 1-2$\sigma$ level.  Therefore, despite the difference in the cloud scheme and retrieved P-T profiles (see Figure \ref{fig:w127_PT_retrievals} vs. Figure \ref{fig:nemesis_triangle}), our main abundance constraints are consistent across the two retrieval codes.

Both our free and equilibrium ATMO retrievals favour super-solar abundances and strong absorption by CO$_{2}$. The 4.5$\mu$m absorption feature in the transmission spectrum is unusually strong, and to the best of our knowledge, no broadband transmission spectrum to date has shown such a strong feature in this wavelength band (e.g. \citealt{2016Natur.529...59S}). 
Both retrievals fit the feature with CO$_2$, which is an indication of high metalicities, as there there is a well-studied sensitivity of CO$_2$ to metallicity (e.g. 
\citealt{2002Icar..155..393L},   
\citealt{2005MNRAS.364..649F}, 
\citealt{2008ApJ...678.1419F}, 
\citealt{2008ApJ...678.1436B}, 
\citealt{2010ApJ...717..496L},
\citealt{2011ApJ...737...15M},
\citealt{2016ApJ...817..149H},
\citealt{2018MNRAS.474.5158G}). 
The free-chemistry model gives a VMR abundance of $-3.83^{+0.27}_{-0.23}$ which is somewhat lower compared to the VMR at 1 mbar from the best-fit equilibrium chemistry model of -2.7$\pm$0.2. 
These values correspond to super solar metallicities, with the chemical equilibrium retrievals finding O and C are 16 and 26 $\times$ solar respectively, with both values well-constrained with 1-$\sigma$ uncertainties of $\sim$0.2 dex (see \ref{tab:arceq}).  With Na, H$_2$O and CO$_2$ all showing signatures of super-solar metallicities, WASP-127b is currently one of the few such cases were the abundances of multiple species can be constrained within the planet's atmosphere.

Our HST/STIS spectrum also shows wavelength-dependent scattering in the optical. We do not see the sharp rise blueward of 5\,600\AA$\,$ that \cite{2018A&A...616A.145C} report from the ground-based NOT data.  Instead, we find a shallower slope, presumably caused by scattering off some kind of condensate/haze material made from small ($<1 \mu m$) particles, which slopes down into the near-infrared.   We note that ground-based optical transmission spectra frequently suffer from differential atmospheric extinction problems at $<0.4\mu m$.  The HST spectrum also contains little evidence of K, and no evidence of Li, which were previously reported to be present from ground-based spectroscopy.

%List all of the planets that have both 3.6 and 4.5 microns and show that this is biggest difference.

To highlight the evidence for CO$_{2}$ in the transmission spectrum, in Figure \ref{fig:w127_spec} we show thre ATMO model atmospheres: our best-fit model from the free-chemistry retrieval; a model with all of the same parameters as the best fit, except the CO$_{2}$ abundance, which is set to zero; and a third model with both CO and CO$_{2}$ abundances set to zero.  The strong absorption feature centred on the 4.5 $\mu$m Spitzer channel disappears in the latter two models. 
With only Spitzer photometry, however, the contribution of CO to the 4.5$\mu$m point complicates the interpretation of the C/O ratio.   
Theoretical models have found that CO should be the dominant carbon-bearing molecule for hydrogen-dominated atmospheres above 1\,000K (e.g. \citealt{2002Icar..155..393L,2016ApJ...817..149H}), and CO has been detected at high resolution in hot Jupiter atmospheres (e.g. \citealt{2010Natur.465.1049S}).
In our equilibrium chemistry retrieval, CO is at least 100$\times$ more abundant than CO$_2$ (see Figure \ref{fig:AMOchem}).  
However, at 4.5$\mu$m the CO$_2$ opacity is much stronger and dominates over the CO contribution even though CO has much higher VMR concentrations. In the free-chemistry retrieval, the CO VMR is not constrained by the data - only an upper-limit to the CO is found, as very high values affect the mean molecular weight, and the data are consistent with no CO contribution. The lack of a CO feature in the WFC3 data further pushes the free-chemistry retrieval to prefer CO$_2$ over CO.  With CO constrained through chemistry in one retrieval and unconstrained in the free case, the C/O ratios obtained are vastly different.  
In the chemical-equilibrium case, a super-solar C/O is found (see Fig. \ref{fig:w127_eq_mcmc}) while in the free-case a sub-solar C/O ratio is found (Fig. \ref{fig:w127_free_mcmc}). This finding highlights the extreme sensitivity and degeneracies of measuring the C/O ratio with a free-chemistry retrieval model, as all major molecular species have to be well constrained by the data. 
For a hot Jupiter such as WASP-127b, we consider a scenario with all of the carbon found in CO$_2$ and little to none in CO to be thermochemically implausible, as no obvious non-equilibrium mechanism would deplete CO by many orders of magnitude while enhancing CO$_2$.  This situation is unlike CH$_4$, where dynamical mixing and vertical quenching can dramatically enhance CH$_4$ (e.g. 
\citealt{2006ApJ...649.1048C,
2011ApJ...737...15M,
2017ApJS..228...20T,
2018ApJ...855L..31D,
2018ApJ...869...28D}). 
%WASP-127b is expected to have an equilibrium temperature of 1\,400K, but on the other hand, our best fit temperature is only 1107$\pm$91K.  There is a CO feature at the $\sim$4.5 micron that may explain the increased opacity there (e.g. \citealt{2019MNRAS.482.4503G}).  
With only one photometric data point at 4.5$\mu$m, it is currently impossible to fully disentangle the contribution of both CO and CO$_2$ in a model-independent way.  Further transmission spectroscopy observations of WASP-127b at high resolution with the James Webb Space Telescope will clarify which is the dominant carbon-bearing molecule in WASP-127b's atmosphere, and allow stronger constraints to be placed on its carbon-to-oxygen ratio.  

\begin{figure}
    \centering
    \includegraphics[width=0.49\textwidth]{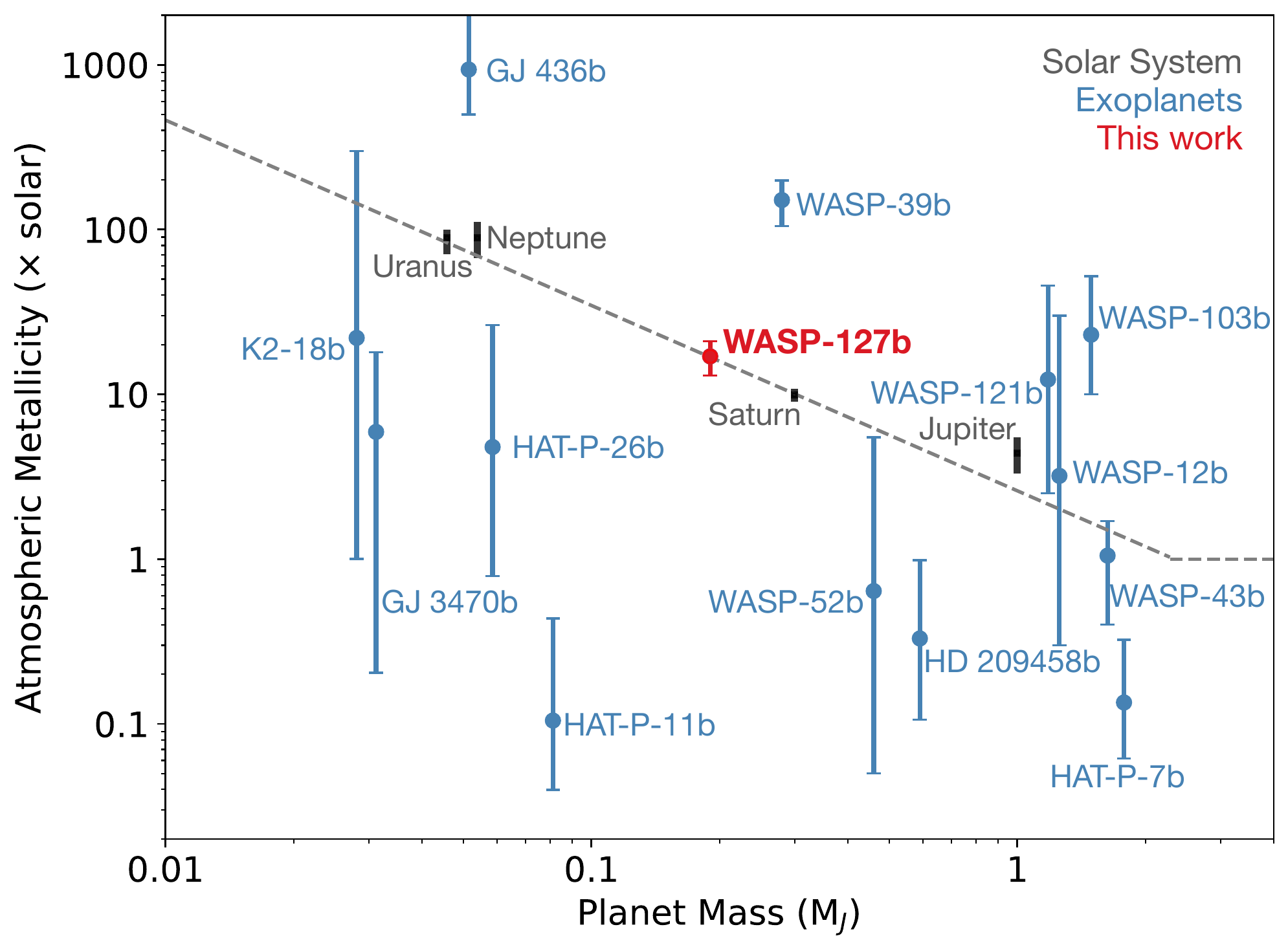}
    \vspace*{-1\baselineskip}          %Use command to remove white space
    \caption{Mass against atmospheric metallicity for the solar system giant planets (grey squares) and measured exoplanets (blue and red dots). The trend seen in the solar system giants is taken to indicate formation via core accretion. Our measurement of WASP-127b's atmospheric metallicity using the equilibrium chemistry oxygen abundance as proxy (red dot) fits squarely on this trend line. However, including the other measured exoplanets there is no clear trend seen throughout the measured exoplanet population.}
    \label{fig:MassMetallicity}
\end{figure}

Although we do not resolve any absorption features of iron hydride (FeH), our best-fit, free-chemistry retrieval from ATMO fits the data better at a confidence of 2.5 $\sigma$ compared to free-chemistry retrievals run without FeH.  The retrieved VMR was -7.56$^{+0.36}_{-0.39}$, which is significantly higher than expected from solar abundances.  Figure \ref{fig:AMOchem} shows that in chemical equilibrium, such high abundances of FeH can be expected deeper in the atmosphere, at pressures of around 60 bars.  There have been several previous reports of FeH in exoplanet atmospheres, including WASP-127b (e.g. \citealt{2020AJ....159....5S} and  \citealt{2020arXiv200509615S}). However,  there may be other, unresolved absorbers in the 0.8 - 1.2 $\mu$m wavelength range, and it is unclear how such high abundances of FeH could be present at such low pressures in WASP-127b's atmosphere.  Since the ATMO chemical equiliubrium model (which folds the FeH abundance into log$_{10}$(Z$_{\rm trace}$/Z$_{\odot}$)) and the free-chemistry model are similarly favoured by the data, with a reduced $\chi^2$ of 0.86 and 0.81 respectively, we do not have enough evidence to definitively detect FeH.   

Our retrieved temperature profiles from ATMO (Figure \ref{fig:w127_PT_retrievals}) agree fairly well with the converged P-T profile for WASP-127b from \cite{Goyal2020arXiv200801856G}.  Of the converged P-T profiles which we ran with an internal temperature of 500 K, and varying heat redistribution factors, the model with $f_{c}$ = 0.25 is most similar to that from our chemical equilibrium retrieval. This suggests that at WASP-127b's terminator, where our transmission signal originates, only 25\% of the incoming stellar irradiation on the planet's dayside is required to match our observed PT profile and abundances in chemical equilibrium. This is as expected from a hot, tidally locked planet which likely re-radiates some of the flux recieved on its dayside before it can be advected around the planet by winds.  Our retrieved NEMESIS P-T profile is noticeably cooler than those from ATMO (Figure \ref{fig:nemesis_pt}), which may be a result of the greater degree of flexibility in the P-T profile fit, and the patchy haze scheme which may allow colder temperatures to mute absorption features, instead of hazes.

To place the abundance measurements in a wider context, we plot the measured atmospheric metallicity from the equilibrium chemistry retrieval against other exoplanet measurements and the solar system giant planets. Figure\,\ref{fig:MassMetallicity} shows the measured metallicity of the solar system giant planets via their abundance of CH$_4$ \citep{wong2004,fletcher2011,karkoschka2011,sromovsky2011}, and exoplanets via predominantly their H$_2$O abundance \citep{kreidberg2014_w43,kreidberg2015,2017ApJ...839L...2B,2019arXiv191105179B,wakeford2018_w39,2019arXiv191007523C,2017Sci...356..628W,2017AJ....153...86M,2019NatAs...3..813B}. The trend of increased atmospheric metallicity with decreasing mass seen for the solar system giant planets is thought to be indicative of core accretion formation. The measurements of WASP-127b place it in the middle of the exoplanet distribution and along the trend shown by the solar system giants. However, for the current exoplanet population, all of which are orbiting much closer to their stars than their solar system mass counterparts, there is no significant trend in the data. 

Overall, we have observed evidence of strong absorption features from several atomic and molecular species, which means that the level of clouds and potential hazes in WASP-127b is not so strong as to prevent the determination of its atmospheric composition.  We find the atmosphere has a super-solar metallicity which is traced by several species (H$_2$O, Na, and CO$_2$).  Assuming chemical-equilibrium, these three species have an average metallicity of 17$\pm$4 $\times$ solar. While there has been considerable spread in retrieved metallicities of exoplanets to date, WASP-127b is in good agreement with the mass-metallicity trend of the solar system (see Fig. \ref{fig:MassMetallicity}). 
This evidence, combined with a long transit duration, means that WASP-127b is the ideal benchmark exoplanet for measuring chemical abundances of exoplanet atmospheres and should be one of the prime targets for James Webb Space Telescope.  In particular, the hint of a large absorption feature around 4.5$\mu$m is strong evidence that future observations of WASP-127b with JWST will be able to measure the abundances of carbon-bearing species in its atmosphere. 

%\begin{table}
%\centering
%\begin{tabular}{lr}
%\hline
%Parameter & Value \\
%\hline
%T$_{\textrm{eff}}$ (K) & 1107$\pm$91 \\
%R$_{P, 1bar}$ (R$_{J}$) & 1.343$\pm$0.005 \\
%Cloud$^{a}$ & 1.16$\pm$0.38 \\
%Haze$^{a}$  & 6.44$\pm$0.35 \\
%H$_{2}$O$^{b}$ & -4.8$\pm$0.3 \\
%CO$_{2}^{b}$   &      -5.7$\pm$0.7 \\
%CO$^{b}$       &    -17.7$\pm$9.7\\
%NH$_{3}$       &    -21.6$\pm$8.1\\
%Na$^{b}$   &   -3.6$\pm$0.5\\
%K$^{b}$     & -4.5$\pm$0.8\\
%Li$^{b}$      &    -20.5$\pm$9.2\\
%\hline
%\end{tabular}
%\caption{Results from ATMO retrieval MPFIT to WASP-127b's transmission spectrum.  $^{a}$The cloud and haze strengths are defined in Section \ref{sec:arc}. $^{b}$The units of chemical composition are natural log of the fractional abundances for each molecule.}
%\label{tab:arcfit}
%\end{table}

\begin{table}
\centering
\begin{tabular}{lrc}
\hline
Parameter & Value & Prior range \\
\hline
$\chi^2_{min}$ & 47.6 \\
N$_{\rm free}$ & 10 \\
N$_{\rm data}$ & 65 \\
\hline
R$_{P, 1bar}$ [R$_{J}$]       & 1.3861$^{+0.0050}_{-0.0058}$            & 1.24 -- 1.52\\
log$_{10}$$(\upkappa_{IR})$     & -2.99$^{+0.85}_{-0.78}$                & -5 -- -0.5  \\
log$_{10}$$(\gamma_{O/IR})$    & -1.92$^{+1.29}_{-1.06}$                & -4 -- 1.5  \\
$\beta$                       &  0.68$^{+0.03}_{-0.05}$                & 0 -- 2   \\
ln($\delta_{\mathrm{haze}})$  &  2.68$^{+0.35}_{-0.29}$                & -10 -- 10 \\
$\alpha_{\mathrm{haze}}$      &  1.50$^{+0.21}_{-0.27}$                & 0 -- 5\\
log$_{10}$(Z$_{\rm trace}$/Z$_{\odot}$)$^{b}$ & -0.17$^{+0.42}_{-0.30}$ &-1 -- 2\\
log$_{10}$(O/O$_{\odot}$)     &  1.21$^{+0.16}_{-0.17}$                 &-1 -- 2\\
log$_{10}$(C/C$_{\odot}$)     &  1.41$^{+0.17}_{-0.18}$                 &-1 -- 2\\
log$_{10}$(Na/Na$_{\odot}$)   &  0.94$^{+0.44}_{-0.58}$                 &-1 -- 2\\
%log$_{10}$(K/K$_{\odot}$)     &  0.66$^{+0.93}_{-1.58}$                &-1 -- 2\\

\hline
\end{tabular}
\caption{Equilibrium chemistry retrieval results fitting to WASP-127b's transmission spectrum.  $^{a}$The cloud and haze strengths are defined in Section \ref{sec:arc}. $^{b}$ log$_{10}$(Z{\rm trace}/Z$_{\odot}$) is not the overall bulk metallicity but contains the abundances for trace species not otherwise individually fit (i.e. all except H, He, O, C, and Na)}.
\label{tab:arceq}
\end{table}
\begin{table}
\centering
\begin{tabular}{lrc}
\hline
Parameter & Value & Prior range\\
\hline
$\chi^2_{min}$ & 42.7 \\
N$_{\rm free}$ & 13 \\
N$_{\rm data}$ & 65 \\
\hline
R$_{P, 1bar}$ [R$_{J}$]       & 1.4505$^{+0.0075}_{-0.0061}$        & 1.24 -- 1.52\\
log$_{10}$$(\upkappa_{IR})$     & -3.41$^{+0.79}_{-0.75}$             & -5 -- -0.5  \\
log$_{10}$$(\gamma_{O/IR})$    & -1.07$^{+0.92}_{-0.94}$             & -4 -- 1.5  \\
$\beta$                       &  -0.092$^{+0.078}_{-0.204}$             & 0 -- 2   \\
ln($\delta_{\mathrm{haze}}$)   & -0.86$^{+0.44}_{-0.63}$             & -10 -- 10 \\
$\alpha_{\mathrm{haze}}$      &  1.06$^{+0.27}_{-0.22}$             & 0 -- 5\\
VMR log$_{10}$(H$_2$O)        & -4.16$^{+0.17}_{-0.22}$             & -14 -- -1.3\\
VMR log$_{10}$(CO$_2$)        & -5.52$^{+0.46}_{-0.39}$             & -14 -- -1.3 \\
VMR log$_{10}$(CO)            & -7.87$^{+2.68}_{-3.04}$             & -14 -- -1.3\\
VMR log$_{10}$(CH$_4$)        & -10.18$^{+1.72}_{-1.75}$            & -14 -- -1.3\\
VMR log$_{10}$(Na)            & -6.99$^{+0.35}_{-0.41}$             & -14 -- -1.3\\
VMR log$_{10}$(K)             & -9.77$^{+1.15}_{-1.26}$             & -14 -- -1.3\\
VMR log$_{10}$(FeH)             & -7.56$^{+0.36}_{-0.39}$             & -14 -- -1.3\\
\hline
\end{tabular}
\caption{Free-chemistry retrieval results fitting to WASP-127b's transmission spectrum. VMR refers to the volume mixing ratio.}
\label{tab:arcfree}
\end{table}

\section*{Acknowledgements}
This work is based on observations made with the NASA/ESA Hubble Space Telescope that were obtained at the Space Telescope Science Institute, which is operated by the Association of Universities for Research in Astronomy, Inc.
Support for this work was provided by NASA through grants under the HST-GO-14619 program from the STScI
This portion of the work is based on observations made with the Spitzer Space Telescope, which is operated by the Jet Propulsion Laboratory, California Institute of Technology under a contract with NASA.  ALC is funded by a UK Science and Technology Facilities (STFC) studentship. This work made use of the python package corner \citep{corner}. J. J. S. thanks the anonymous reviewer for productive comments.

%The Acknowledgements section is not numbered. Here you can thank helpful
%colleagues, acknowledge funding agencies, telescopes and facilities used etc.

\section{Data Availability}
Raw HST data frames are publicly available online at the Mikulski Archive for Space
Telescopes (MAST; https://archive.stsci.edu).  Raw Spitzer data frames are publicly available at the NASA/IPAC Infrared Science Archive (IRSA; https://sha.ipac.caltech.edu/applications/Spitzer/SHA/)

%%%%%%%%%%%%%%%%%%%%%%%%%%%%%%%%%%%%%%%%%%%%%%%%%%



\bibliographystyle{mnras}
\bibliography{bib3} % if your bibtex file is called example.bib

%%%%%%%%%%%%%%%%%%%%%%%%%%%%%%%%%%%%%%%%%%%%%%%%%%

%%%%%%%%%%%%%%%%% APPENDICES %%%%%%%%%%%%%%%%%%%%%

\appendix

\section{TESS Photometry}

\begin{figure*}
    \centering
    \includegraphics[width=\textwidth]{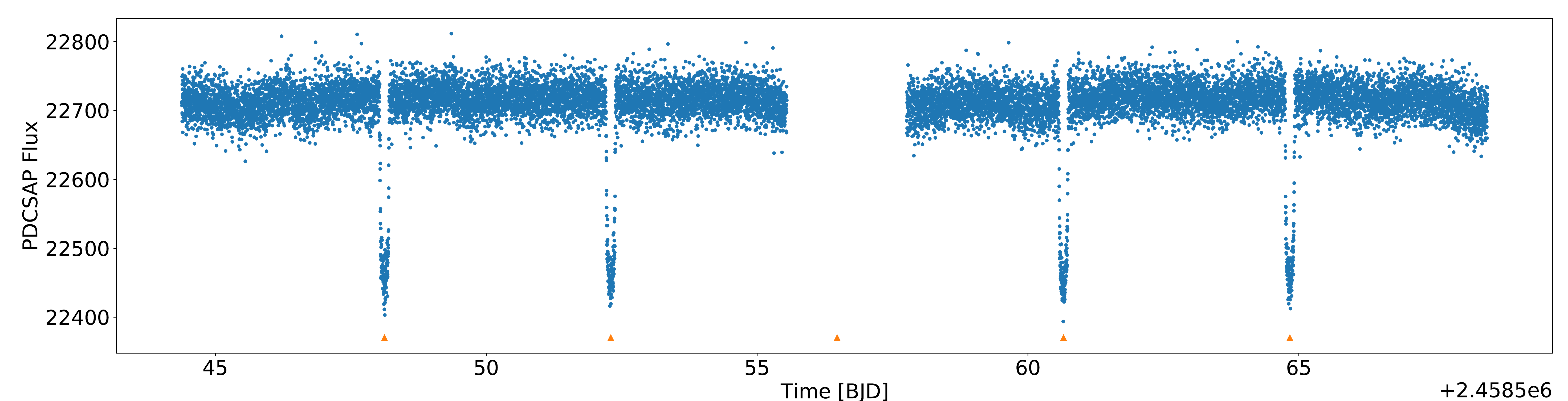}
    \caption{Blue points are TESS photometry of WASP-127, orange triangles show expected mid-transit times of WASP-127b.}
    \label{fig:TESS_full}
\end{figure*}

\begin{figure}
    \centering
    \includegraphics[width=\columnwidth]{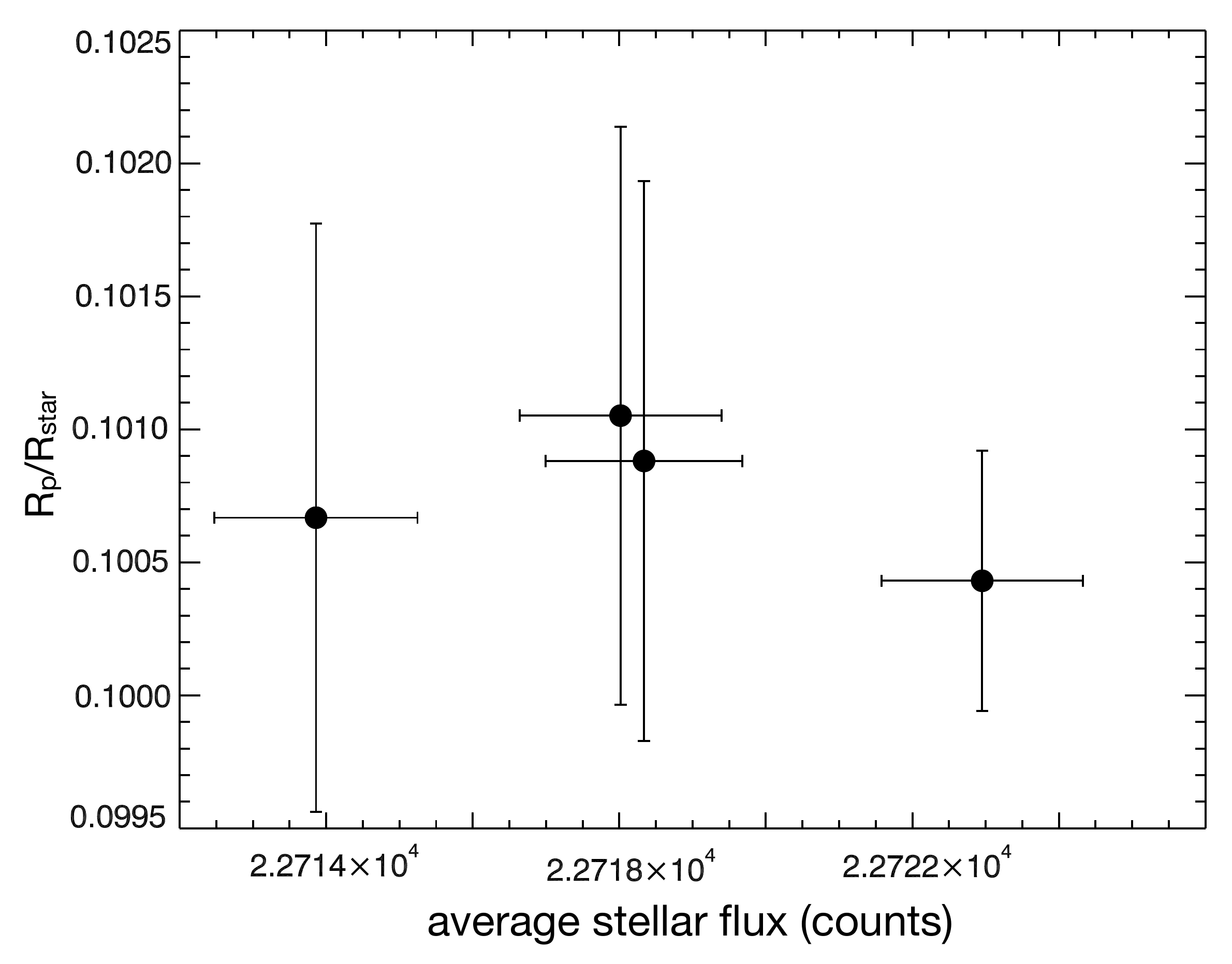}
    \caption{Measured planet-to-star radius ratios from four TESS transit events against average stellar flux.}
    \label{fig:TESS_tdepths}
\end{figure}

\section{Light curve fitting corner plots}
\begin{figure*}
    \centering
    \includegraphics[width=0.6\textwidth]{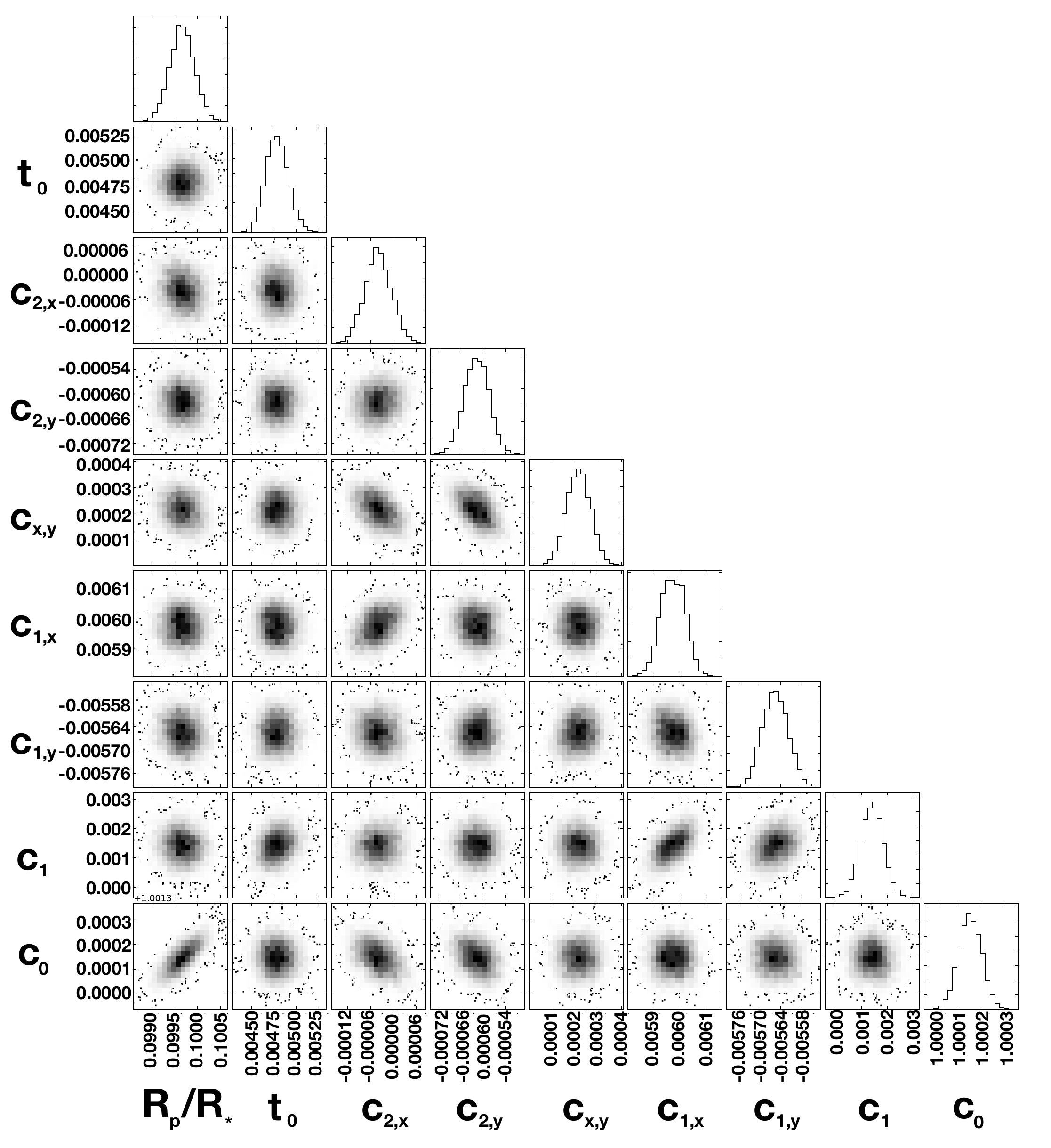}
    \vspace*{-1\baselineskip}          %Use command to remove white space
    \caption{Posterior distributions for lightcurve MCMC fits for WASP-127b, using Spitzer/IRAC's 3.6$\mu$m channel.}
    \label{fig:w127_ch1triangle}
\end{figure*}

\begin{figure*}
    \centering
    \includegraphics[width=0.6\textwidth]{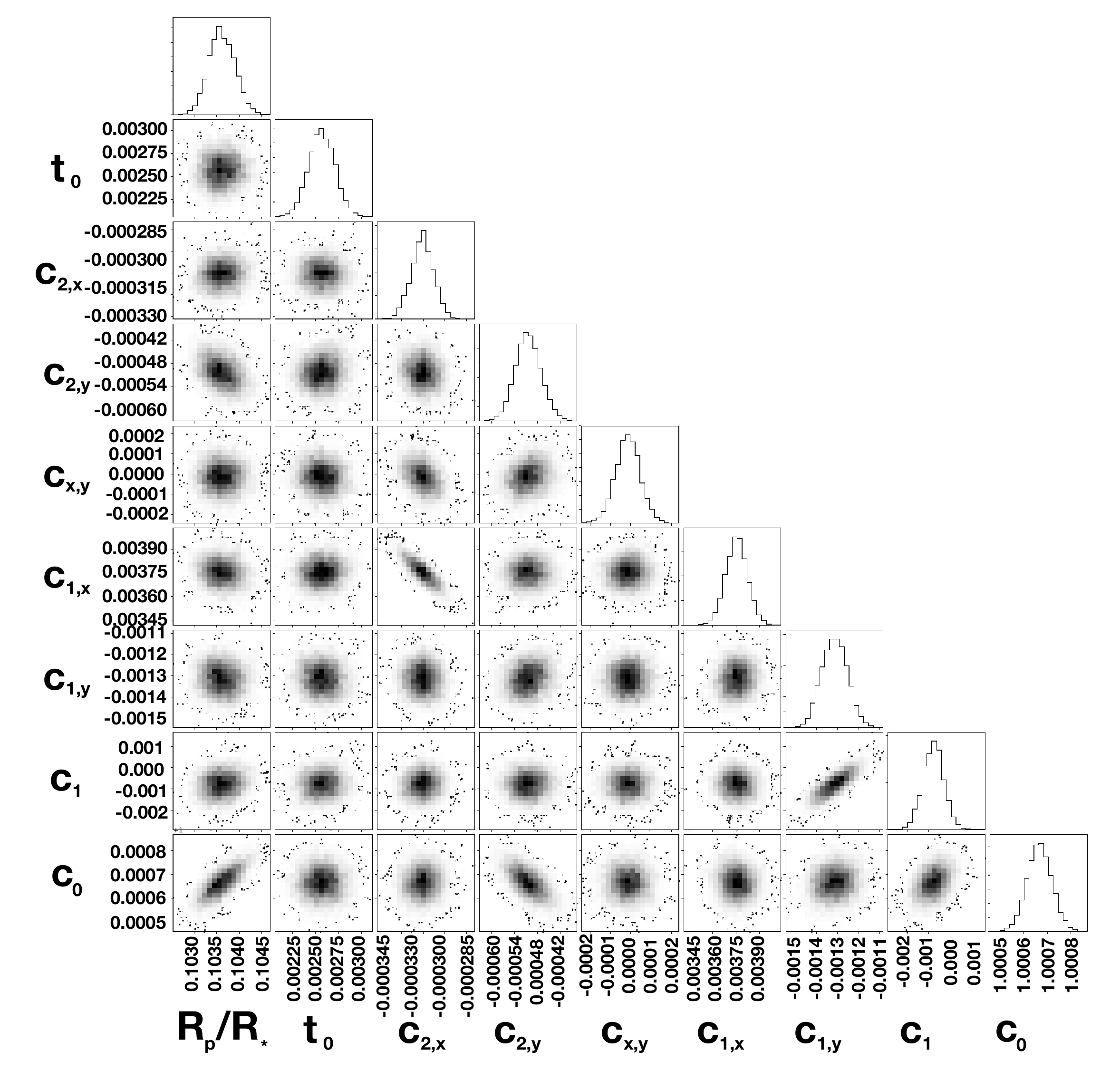}
    \vspace*{-1\baselineskip}          %Use command to remove white space
    \caption{Posterior distributions for lightcurve MCMC fits for WASP-127b, using Spitzer/IRAC's 4.5$\mu$m channel.}
    \label{fig:w127_ch2triangle}
\end{figure*}

\begin{figure*}
    \centering
    \includegraphics[width=0.6\textwidth]{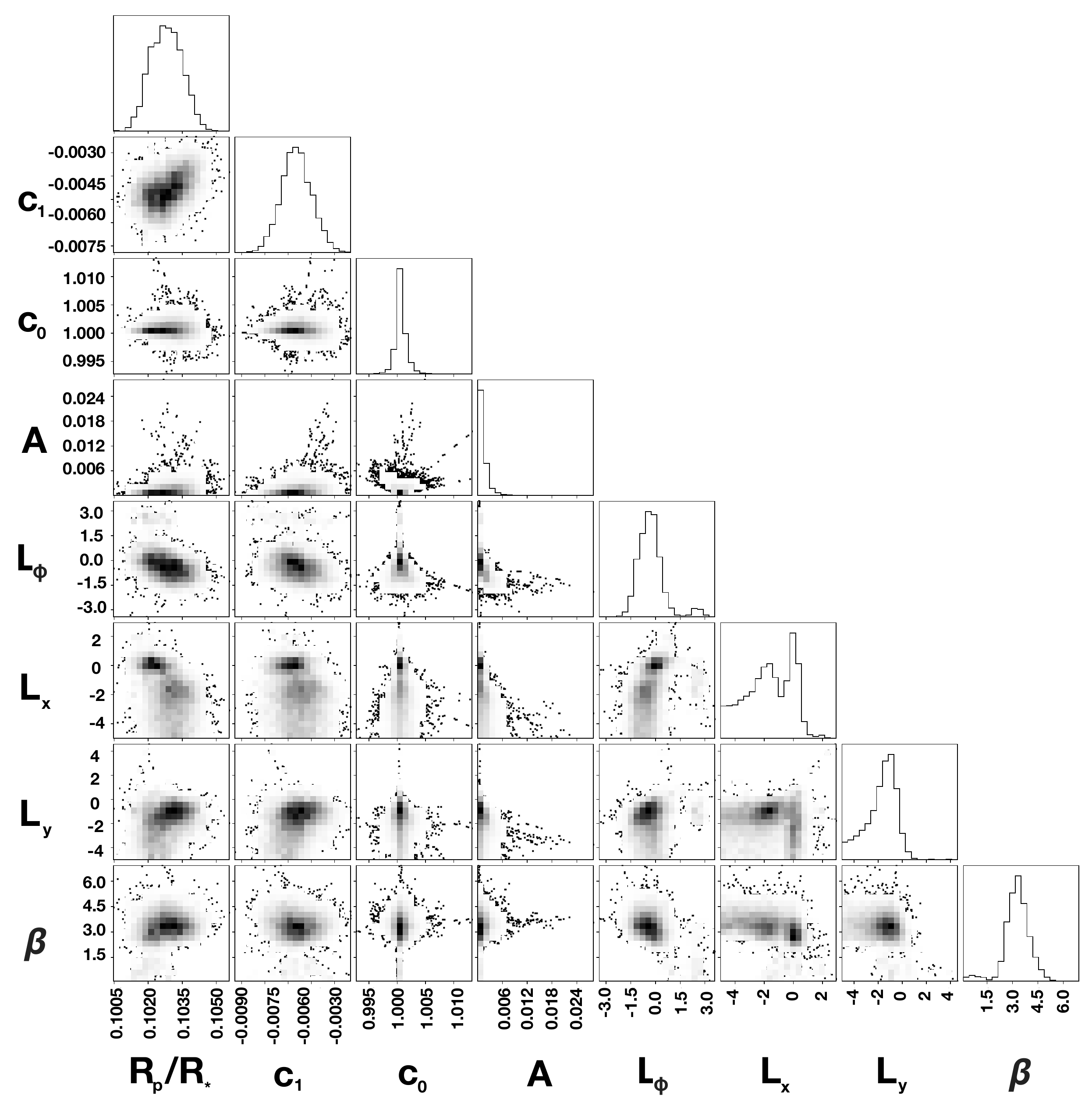}
    \vspace*{-1\baselineskip}          %Use command to remove white space
    \caption{Posterior distributions for white lightcurve MCMC fit for WASP-127b, using HST/STIS+G430L.}
    \label{fig:w127_g430triangle_white}
\end{figure*}

\begin{figure*}
    \centering
    \includegraphics[width=0.6\textwidth]{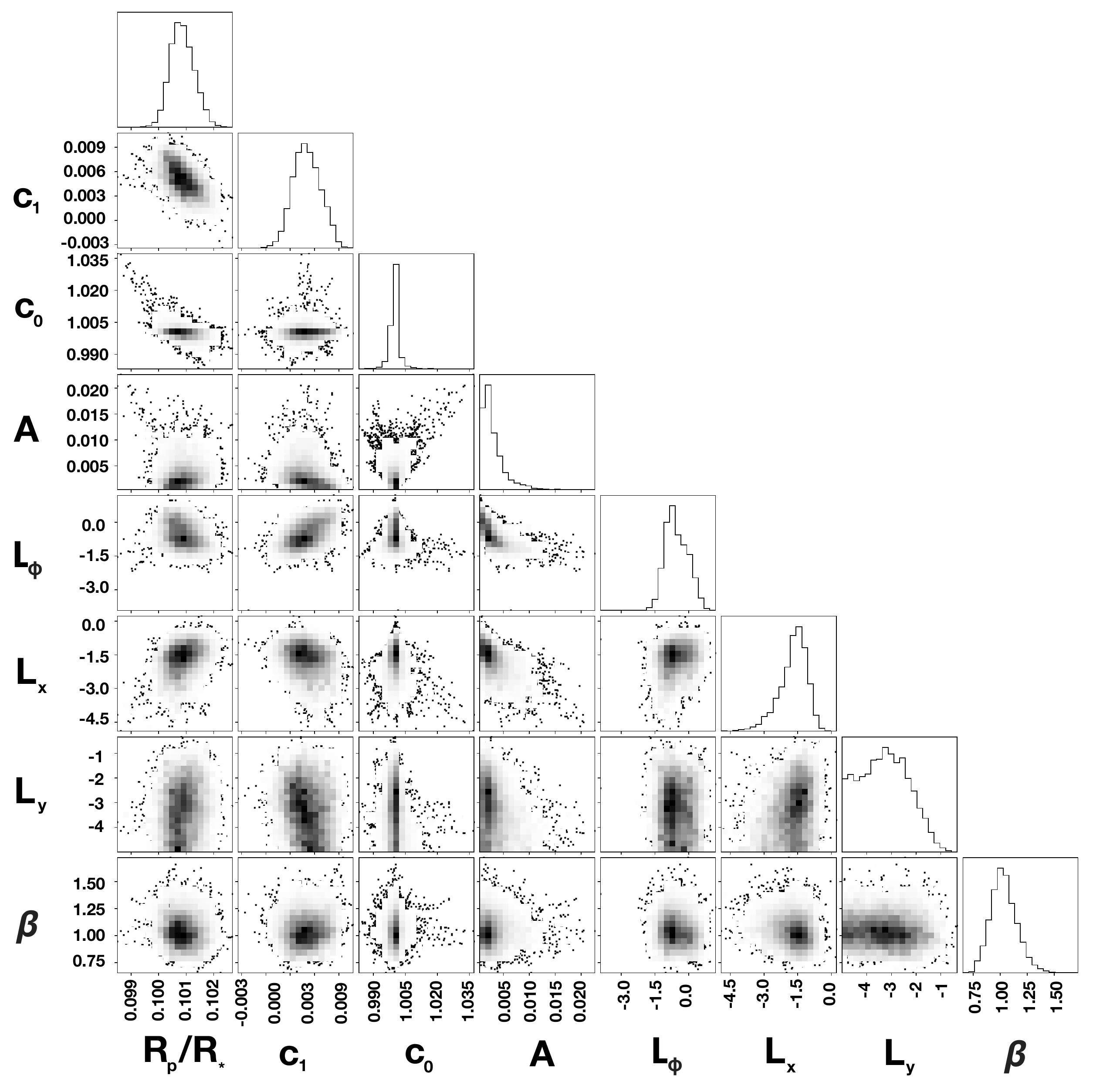}
    \vspace*{-1\baselineskip}          %Use command to remove white space
    \caption{Posterior distributions for white lightcurve MCMC fit for WASP-127b, using HST/STIS+G750L.}
    \label{fig:w127_g750triangle_white}
\end{figure*}

\begin{figure*}
    \centering
    \includegraphics[width=0.6\textwidth]{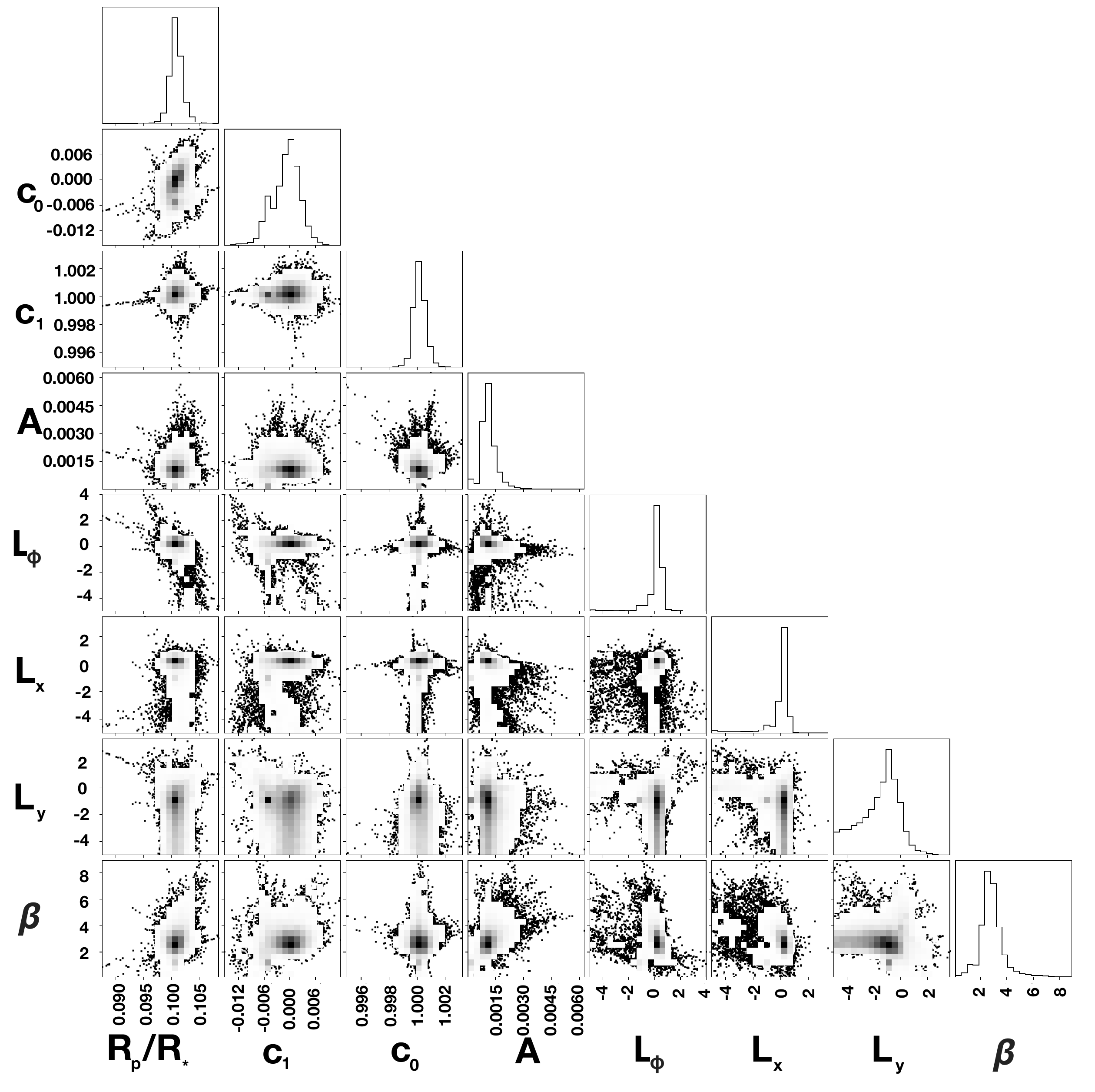}
    \vspace*{-1\baselineskip}          %Use command to remove white space
    \caption{Posterior distributions for white lightcurve MCMC fit for WASP-127b, using HST/WFC3+G141.}
    \label{fig:w127_wfc3triangle_white}
\end{figure*}

\begin{figure*}
    \centering
    \includegraphics[width=0.6\textwidth]{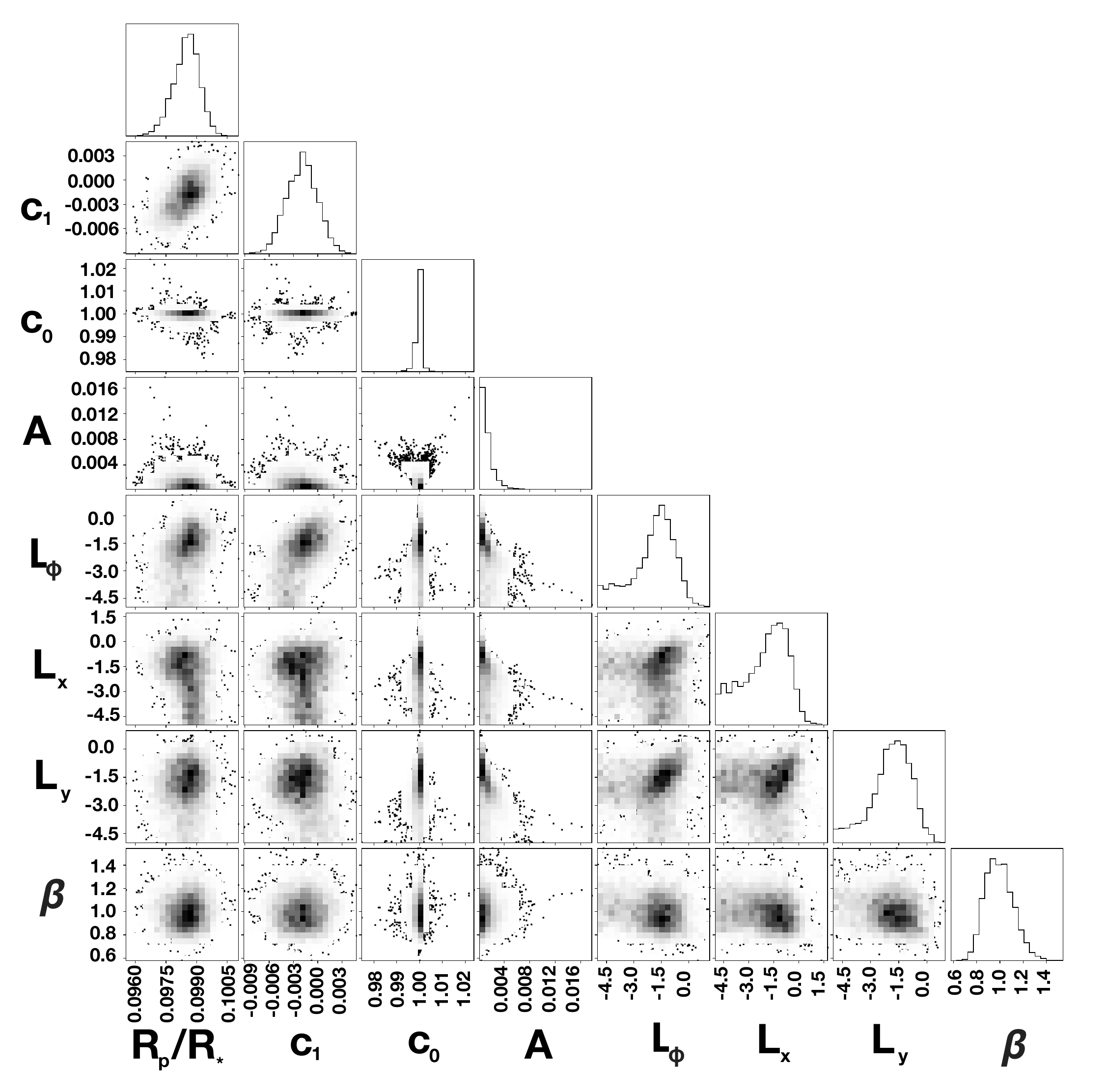}
    \vspace*{-1\baselineskip}          %Use command to remove white space
    \caption{Typical posterior distributions for spectroscopic lightcurve MCMC fits for WASP-127b, using HST/WFC3+G141.}
    \label{fig:w127_wfc3triangle}
\end{figure*}

\begin{figure*}
    \centering
    \includegraphics[width=0.6\textwidth]{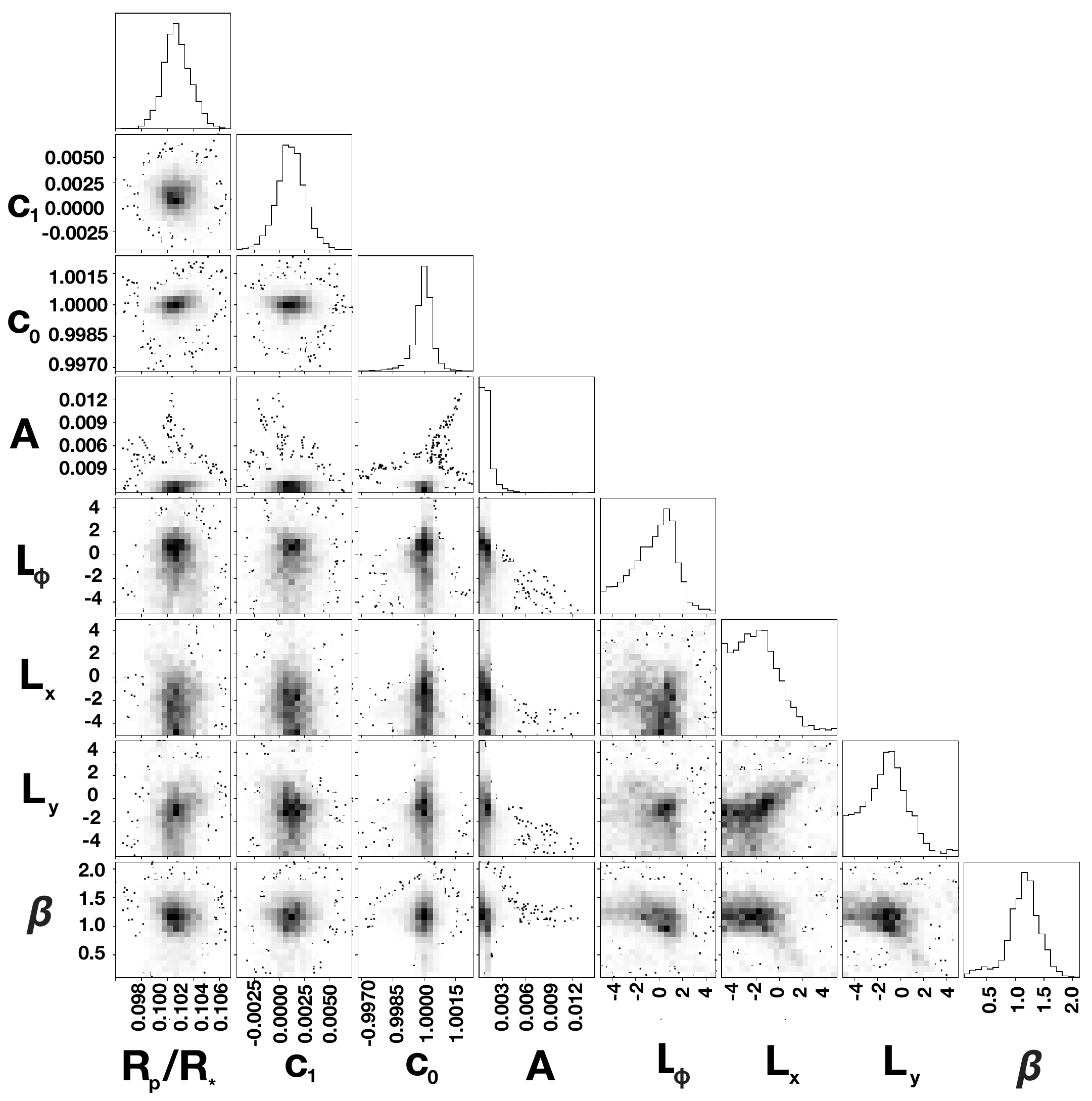}
    \vspace*{-1\baselineskip}          %Use command to remove white space
    \caption{Typical posterior distributions for spectroscopic lightcurve MCMC fits for WASP-127b, using HST/STIS+G430L.}
    \label{fig:w127_g430triangle}
\end{figure*}

\begin{figure*}
    \centering
    \includegraphics[width=0.6\textwidth]{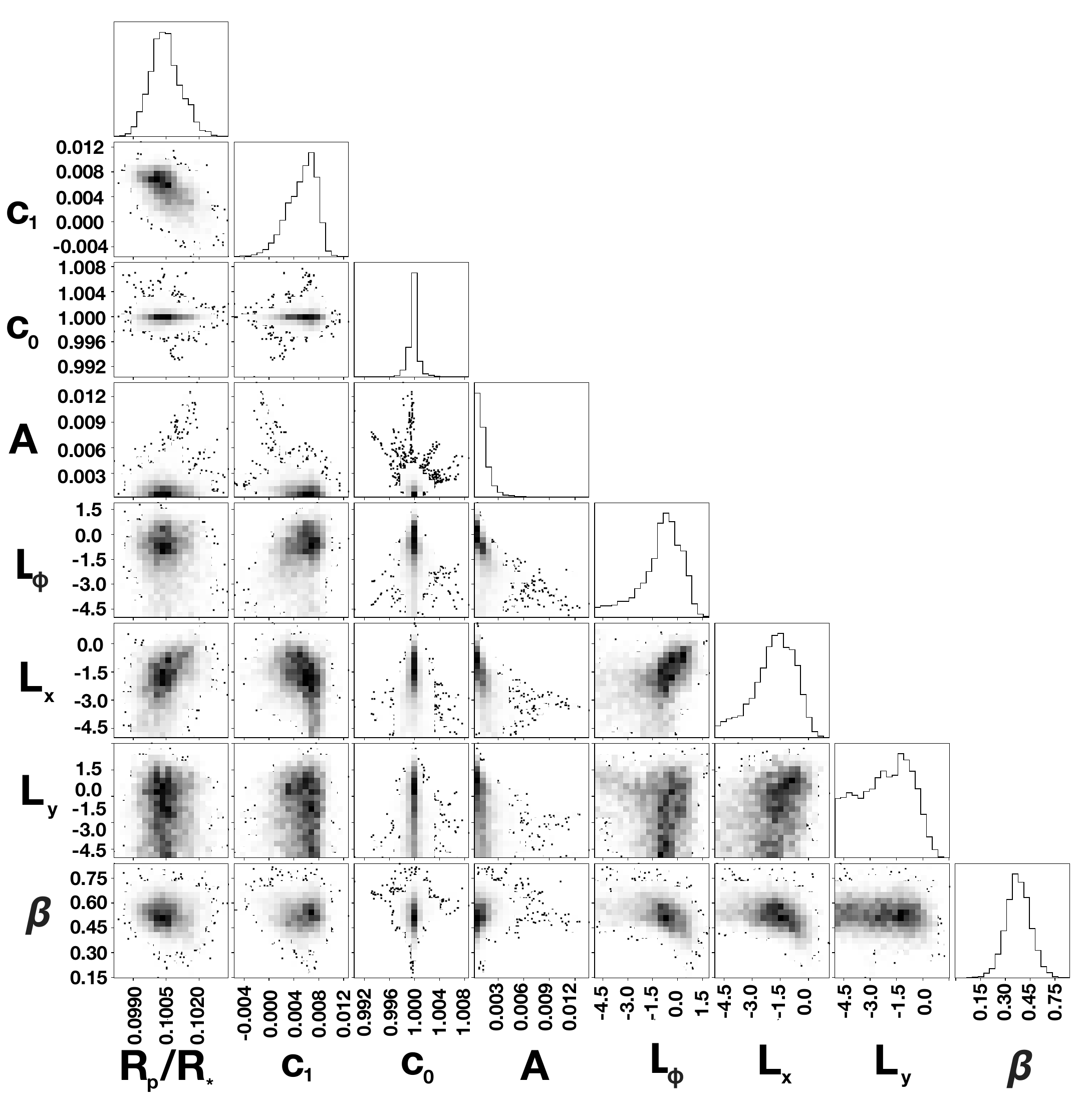}
    \vspace*{-1\baselineskip}          %Use command to remove white space
    \caption{Typical posterior distributions for spectroscopic lightcurve MCMC fits for WASP-127b, using HST/STIS+G750L.}
    \label{fig:w127_g750triangle}
\end{figure*}


%%%%%%%%%%%%%%%%%%%%%%%%%%%%%%%%%%%%%%%%%%%%%%%%%%

% Don't change these lines
\bsp	% typesetting comment
\label{lastpage}
\end{document}